\documentclass[%
    a4paper,
    10pt,
    DIV=11,
    BCOR=0mm,
    twoside=semi,
    headings=small
]{scrartcl}
\usepackage{scrpage2}

\addtokomafont{title}{\normalfont \bfseries}
\addtokomafont{section}{\normalfont \bfseries}
\addtokomafont{subsection}{\normalfont \bfseries}
\addtokomafont{subsubsection}{\normalfont \bfseries}
\addtokomafont{paragraph}{\normalfont \bfseries}
\newcommand{\quoteparagraph}[1]{\noindent{\normalsize \bfseries #1}\hspace{1em}}

\usepackage[utf8]{inputenc}           
\usepackage[font=scriptsize,labelfont=bf]{caption} 
\usepackage{amsmath, amsthm, amssymb} 
\usepackage[english]{babel}           	
\usepackage{bibgerm}  					
\usepackage{graphicx,subfigure}       
\usepackage{hyperref}                 

\newcommand{\E}{\ensuremath{\mathbb{E}}}
\newcommand{\Pe}{\ensuremath{\mathbb{P}}}

\newtheorem{Definition}{Definition}
\newtheorem{Lemma}{Lemma}

\newtheorem{Observation}{Observation} 

\usepackage{bookmark}
\hypersetup{
  pdftitle    = {Survey on log-normally distributed market-technical trend data},
  pdfauthor   = {Ren\'e Kempen and Stanislaus Maier-Paape},
  pdfkeywords = {log-normal, market-technical trend, MinMax-process, trend
  statistics, market analysis, empirical distribution},
  pdfborder={0 0 0.5} }

\author{
\normalsize \textsc{Ren\'e Kempen}\\[-0.2em]
    \small \textit{Institut f\"ur Mathematik, RWTH Aachen,}\\[-0.5em]
    \small \textit{Templergraben 55, D-52052 Aachen, Germany}\\[-0.5em]
    \small
    \href{mailto:kempen@instmath.rwth-aachen.de}{kempen@instmath.rwth-aachen.de}
    \\
  \\[-0.75em]
  \normalsize \textsc{Stanislaus Maier-Paape}\\[-0.2em]
    \small \textit{Institut f\"ur Mathematik, RWTH Aachen,}\\[-0.5em]
    \small \textit{Templergraben 55, D-52052 Aachen, Germany}\\[-0.5em]
    \small \href{mailto:maier@instmath.rwth-aachen.de}{maier@instmath.rwth-aachen.de}
    } \date{
  \vspace{0.25em}
  \normalsize\today
  \vspace{-1cm}
}
\title{
  \vspace{-2cm}
  \Large Survey on log-normally distributed market-technical trend data
}

\clearscrheadfoot
 \lehead{\pagemark}                   
\rohead{\pagemark}                   
\begin{document}

\maketitle

\begin{quote}
  \small
  \quoteparagraph{Abstract}
  In this survey, a short introduction in the recent discovery of log-normally
distributed market-technical trend data will be given. The results of the
statistical evaluation of typical market-technical trend variables will be
presented. It will be shown that the log-normal assumption fits better to empirical trend data than to daily returns of stock prices.
This enables to mathematically evaluate trading systems depending on
such variables. In this manner, a basic approach to an anti cyclic
trading system will be given as an example.
\quoteparagraph{Keywords} log-normal, market-technical trend, MinMax-process,
  trend statistics, market analysis, empirical distribution, quantitative
  finance
\end{quote}

\section{Introduction}
The concept of a trend has been fundamental in the field of technical analysis
since Charles H. Dow introduced it in the late 19th century. Following Rhea
\cite{RHEA}, Dow said e.g. concerning the characterization of
\textit{up-trends}:\par
\textit{Successive rallies penetrating preceding high points, with ensuing
declines terminating above preceding low points, offer a bullish indication.}
\par
In Figure \ref{fig:trend}.(a) an example of the inverse situation is given, i.e.
a \textit{down-trend} in a historical setup like Dow used it. Although this is
so far ``just'' a geometrical idea and clearly not precise at all, it is widely
accepted among many market participants. Therefore, this geometric idea is
fixed by the following market-technical definition of a Dow-trend as which it is
also used in this article:
\begin{Definition}[market-technical trend or Dow-trend]\label{trend}
A market is in \textit{up/down-trend} if and only if (at least) the two
last relevant \textit{lows} (denoted by P$1$ and P$3$ in an up-trend) and
\textit{highs} (denoted by P$2$ in an up-trend) are monotonically
increasing/decreasing (see Figure \ref{fig:trend}.(b)).
Otherwise, the market is temporarily \textit{trendless}.
In case of an up-trend the phase between a low and the next high is called the
\textit{movement}. In the same manner, the phase between a high and the
following low is called the \textit{correction}. In case of a
down-trend, movement and correction are defined in the exact opposite way.
\begin{figure}[h]
  \centering
  \subfigure[Example of a down-trend in a historical setup like Dow used
it (freely adapted from Russel,\cite{RUSSEL}).]{
	\includegraphics[keepaspectratio=true,width=0.45\linewidth]{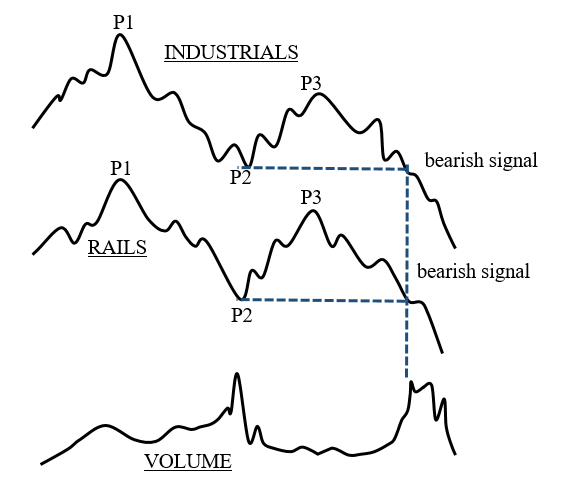}
	}
	\subfigure[Sketch of an up-trend in the sense of Dow]{
	\includegraphics[keepaspectratio=true,width=0.45\linewidth]{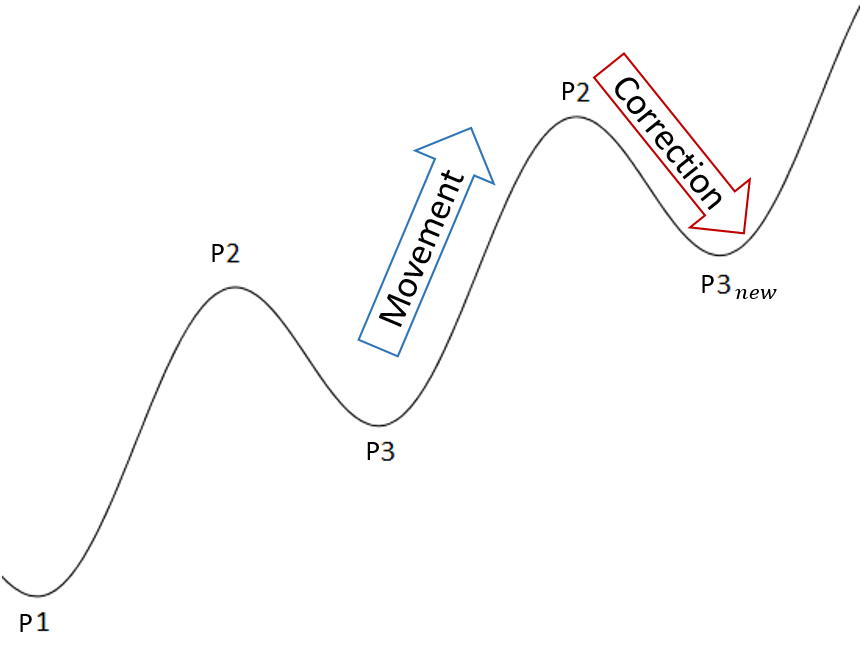}
	}
\caption{}
\label{fig:trend}
\end{figure}

\end{Definition}
It is the authors desire to analyze these Dow-trends as they occur in real world
markets within a statistical framework. In order to do so, however, a
mathematical exact method on how to determine relevant lows and highs of price
data is needed.\par
While the task to detect the extreme points in Figure \ref{fig:trend}.(b)) is
trivial, it is not as easy when a real price chart is considered (see Figure
\ref{fig:minmax}). This can be explained by the continuous price fluctuations
which make the extreme points $P1-P3$ not obvious to detect. The issue of
detection is rooted in the subjectivity of the distinction between usual fluctuations and new extreme points. So to say: the
significance of an extreme point has to be evaluated in an algorithmic way to make automatic detection possible.
Therefore, we review in section \ref{sec:1} the framework which is necessary for automatic trend-detection going back to Maier-Paape \cite{SMP_automatic123}.
The trend-detection in turn is rooted in automatic recognition of relevant
minima and maxima in charts.
With that at hand empirical studies of trend data can be made as
seen in \cite{EMP123} and \cite{LEADLAG}. On the one hand, in \cite{EMP123}
Hafizogullari, Maier-Paape and Platen have collected several statistics on the performance of Dow-trends. On the other
hand, in \cite{LEADLAG} Maier-Paape and Platen constructed a geometrical method
on how to detect lead and lag when two markets are directly compared -- also
based on the automatic detection of relevant highs and lows.\par
In this article, however, we want to pursue a different path. We are interested
in several specific trend data such as the retracement and the relative
movement and correction. Since these trend data are central for the whole paper,
we here give a precise definition.\par
The first random variable describing trend data which will be important in the
following is the \textit{retracement} denoted by $X$. The retracement is defined as the
size-ratio of the correction and the previous movement, i.e.
\begin{equation}\label{eqn_retr}
X:=\frac{Correction}{Movement}.
\end{equation}
Hence, in case of an up-trend this is given by: 
\begin{equation*}
X=\frac{P2-{P3}_{\text{new}}}{P2-P3}.
\end{equation*}
Another common random variable is the
\textit{relative movement} which for an up-trend is defined by the ratio of the movement and the last low, i.e.
\begin{equation}\label{eqn_bew}
M:=\frac{Movement}{last\,Low}=\frac{P2-P3}{P3}
\end{equation}
and the \textit{relative correction} which for an up-trend is defined as
the ratio of the correction and the last high, i.e.
\begin{equation}\label{eqn_kor}
C:=\frac{Correction}{last\,High}=\frac{P2-P3_{\text{new}}}{P2}.
\end{equation}
In case of a down-trend all situations are mirrored, such that:
\begin{equation*}
X=\frac{{P3}_{\text{new}}-P2}{P3-P2},\quad
M:=\frac{Movement}{last\, High}=\frac{P3-P2}{P3},\quad
C:=\frac{Correction}{last\,Low}=\frac{P3_{\text{new}}-P2}{P2}.
\end{equation*}
The main scope of this survey is to collect and extend results on how the above
defined trend variables (plus several other related ones) can be statistically
modeled. By doing
this the log-normal distribution occurs frequently. Evidently, the log-normal
distribution is very well known in the field of finance. We start off, in
section \ref{sec:2}, by giving a mathematical model of the retracement during
Dow-trends and the delay of their recognition. Furthermore, the duration of the
retracements and their joint distribution with the retracement will be
evaluated. The results on relative movements and relative corrections during
trends will be presented in section \ref{sec:3}. In section \ref{sec:4} it will
be demonstrated how the so far gained distributions of trend variables may be used to model
trading systems mathematically.\par
It will be evident, that the described trend data mostly fit very well to the
log-normal distribution model, although there are significant aberrations for the
duration of retracements (see subsection \ref{sec:delay}). In the past there
have been several attempts to match the log-normal distribution model to the
evolution of stock prices. It already started in 1900 with the PhD Thesis of
Louis Bachelier (\cite{BACHELIER}) and the approach to use the geometric
Brownian motion to describe the evolution of stock prices. This yields
log-normally distributed daily returns of stock prices. Nowadays, the geometric
Brownian motion is widely used to model stock prices (see \cite{HULL})
especially as part of the Black-Scholes model (\cite{BLACKSCHOLES}).
Nevertheless, it has to be noted that empirical studies have shown that the
log-normal distribution model does not fit perfectly to daily returns (e.g. see
Fama \cite{FAMA63}, \cite{FAMA} who refers to Mandelbrot
\cite{MANDELBROT}).\par Overall, we got the
impression that most of the trend data we describe here fit better to the
log-normal distribution model than daily returns of stock prices, although it would be beyond the scope of this paper to do a formal comparison.
In any case, the here observed empirical facts of trend data contribute to a
complete new understanding of financial markets. Furthermore, with the
relatively easy calculations based on the link of the log-normal distribution
model to the normal distribution, actually complex market processes can now
be discussed mathematically (e.g. with the truncated bivariate moments, see
Lemma 1.21 in \cite{MT_Kempen}).

\section{Detection of Dow-trends}\label{sec:1}

The issue of automatic trend-detection has been addressed by Maier-Paape
\cite{SMP_automatic123}. Clearly, the detection of relevant extreme points is a
necessary step to detect Dow-trends. Fortunately, the
algorithm introduced by Maier-Paape allows automatic detection of relevant
extreme points in any market since it constructs so called
\textit{MinMax-processes}.

\begin{figure}[h]
  \centering	
\subfigure[Scaling=$1$]{\includegraphics[keepaspectratio=false,width=0.5\linewidth,height=0.4\linewidth]{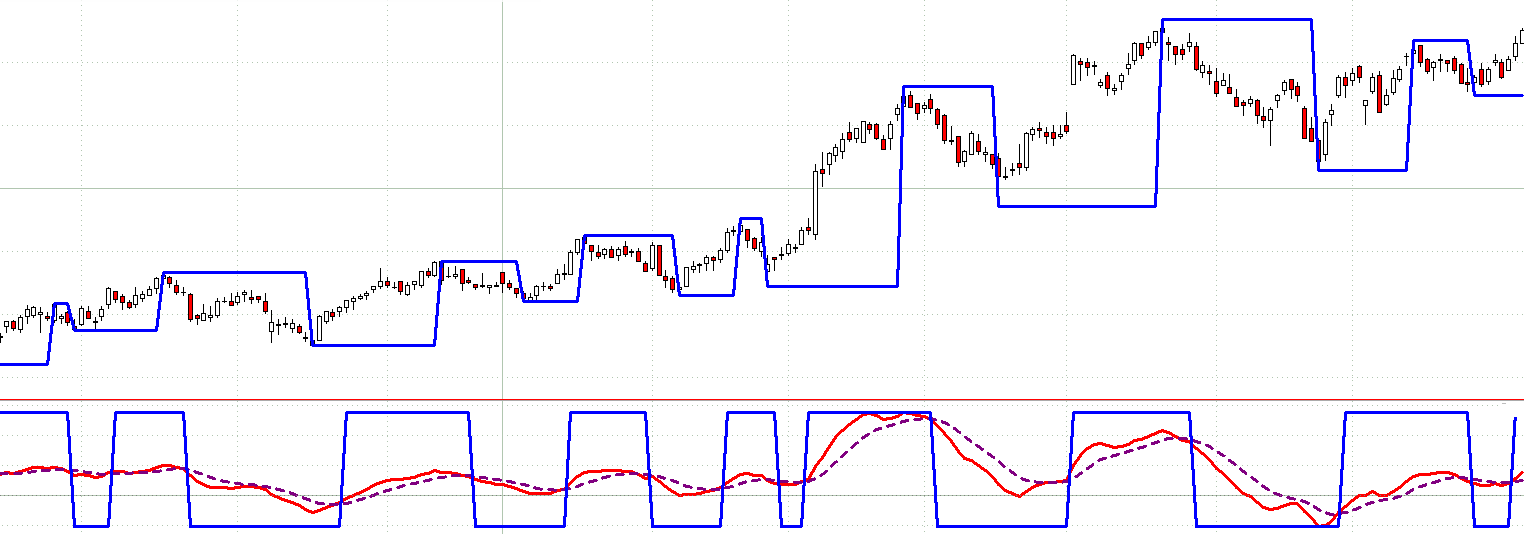}
}\subfigure[Scaling=$3$]{\includegraphics[keepaspectratio=false,width=0.5\linewidth,height=0.4\linewidth]{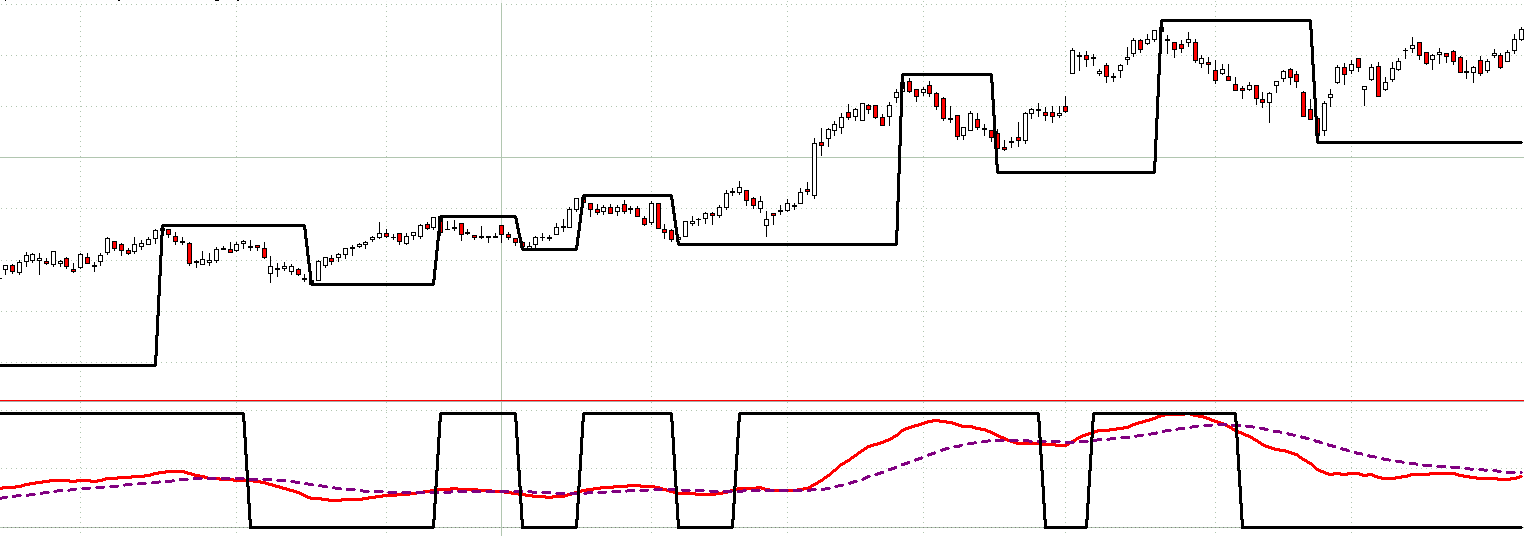}
}
\caption{Daily Chart of the Adidas stock between September 2012 and August 2013
with MinMax-process by Maier-Paape (top) based on the integral
MACD SAR-process (bottom) with two different scalings $1$ and $3$. The lines in the charts indicate the last detected extremum.}
\label{fig:minmax}
\end{figure}

\begin{Definition}[MinMax-process, Definition $2.6$
in \cite{SMP_automatic123}] An alternating series of (relevant) highs and lows
in a given chart is called a \textit{MinMax-process}.
\end{Definition}
 In Figure \ref{fig:minmax} two automatically constructed MinMax-processes
 are visualized by the corresponding indicator line. The construction is based
 on SAR-processes (\underline{s}top \underline{a}nd \underline{r}everse).
\begin{Definition}[SAR-Process]
An indicator is called a \textit{SAR-process} if it can only take two
values (e.g. $-1$ and $1$ which are considered to indicate a $down$ and an $up$
move of the market respectively).
\end{Definition}
Generally speaking, Maier-Paape's algorithm looks for relevant highs when the
SAR-process indicates an up move and searches for relevant lows when the SAR-process
indicates a down move. Thus, the relevant extrema are "fixed" when the SAR-process changes sign. By choosing a specific SAR-process one can affect the
sensitivity of the detection while the actual detection
algorithm works objectively without the need of any further parameter. For more
information see \cite{SMP_automatic123}. Maier-Paape also explains how to handle
specific exceptional situations, e.g. when a new significant low suddenly
appears although the SAR-process is still indicating an up move.
\par
\begin{figure}[h]
  \centering
	\includegraphics[keepaspectratio=true,width=0.6\linewidth]{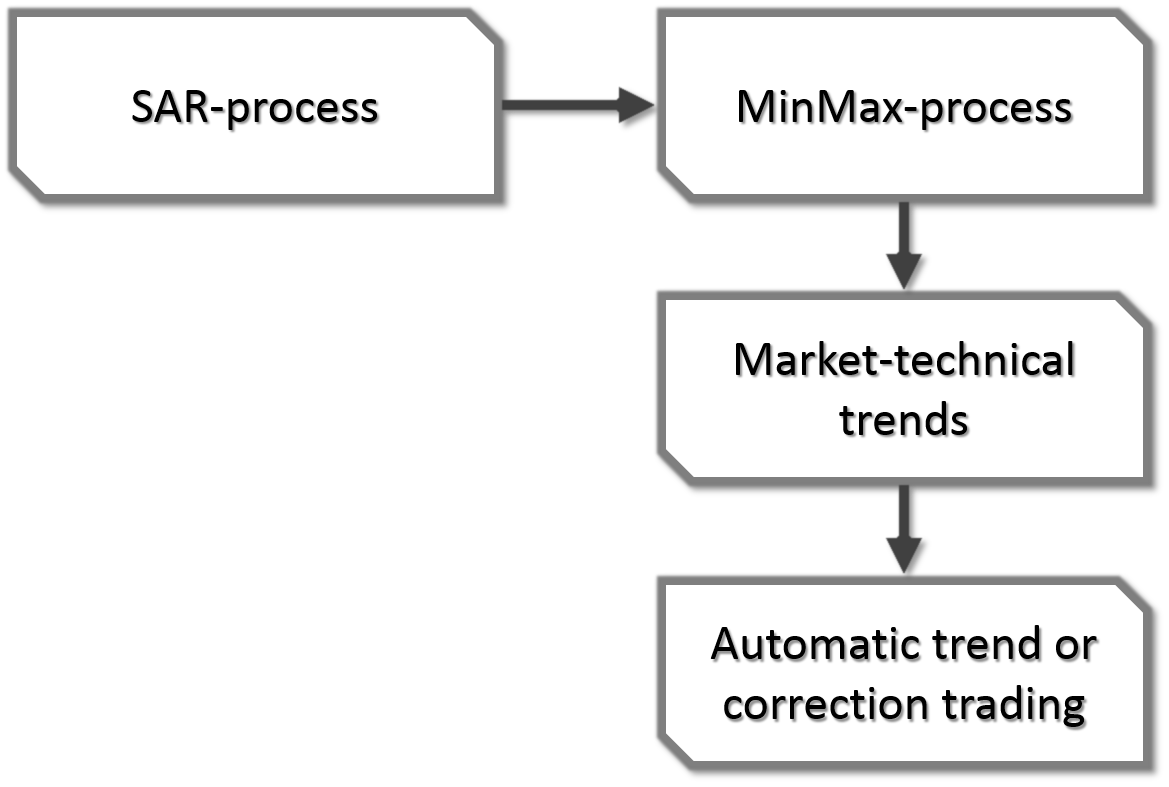}
\caption{General concept of the automatic detection of Dow-trends.}
\label{fig:minmax_struc}
\end{figure}

It is shown by Theorem $2.13$ in \cite{SMP_automatic123} that for any
combination of SAR-process and market there exists such a MinMax-process which can be calculated ``in real
time'' by the algorithm of Maier-Paape. Based on any
MinMax-process in turn it is easy to detect
market-technical trends as defined in Definition \ref{trend} and then use this
information for automatic trading systems as outlined in Figure
\ref{fig:minmax_struc}.\par 
Calculating the MinMax-process ``in real time'' means that as time passes and
the chart gets more and more candles, the extrema of the MinMax-process are
constructed one by another. Besides the most recent extremum which is being
searched for, all extrema found earlier are fixed from the moment of their
detection, i.e. when the SAR-process changed sign.\par Thus, applying the
algorithm in real time also reveals some time \textit{delay} in
detection. Obviously, the algorithm can not predict the future progress of the
chart it is applied to. Consequently, some delay is needed indeed to evaluate
the significance of a possible new extreme value. This circumstance is crucial
when considering automatic trading systems based on market-technical trends.
Therefore, it also has to affect any mathematical model of such a trading
system. An approach to this issue can be made by considering the delay as an
inevitable slippage. This means, not the time aspect of the delay but more
likely the effect it has to the entry or exit price in any market-technical
trading system will be evaluated. In particular, the absolute value of the delay $d_{abs}$ is given by
\begin{equation}\label{def_delay}
d_{abs}=|P[0]-C[0]|
\end{equation}
with $P[0]$ indicating the last detected extreme value and $C[0]$ the close
value of the current bar when this extreme value got detected.
\par For this article the
MinMax-process together with the \textit{integral MACD SAR-process}
(\underline{m}oving \underline{a}verage \underline{c}onvergence
\underline{d}ivergence, see p. $166$ in \cite{MACD}) was used.
The integral MACD SAR (Definition $2.2$ in \cite{SMP_automatic123}) basically is
a normal MACD SAR which in turn indicates an up move if the so called
\textit{MACD line} is above the so called \textit{signal line}. Otherwise, it indicates a
down move. The MACD line is given by the difference of a fast
and a slow (exponential) moving average. The signal line then is an
(exponential) moving average of the MACD line.\par
Consequently, the MACD usually takes three
parameters for the fast, slow and signal line (standard values are: fast=12,
slow=26, signal=9). To reduce the number of needed parameters from three to one
\textit{scaling parameter} only, the ratios of the standard parameters are
fixed and consequently scaled by the scaling parameter. In particular, a MACD
with scaling parameter $2$ denotes a usual MACD with the parameters $(24/52/18)$. This way, the sensitivity of the
MinMax-process solely corresponds to one scaling parameter (see Figure
\ref{fig:minmax}).\par
For a given MinMax-process it is easy to decide the start and end of a
market-technical trend for the candle the trend is initialized and ends in,
respectively.
The computation of several trend variables such as
the retracement is then obvious.\par
The automatic detection of Dow-trends and in particular the possibility to
deduce a MinMax-process out of any market given by candle data enables to create a large dataset of
empirical trend variables. 
The model will be based on empirical data acquired by the MinMax-process based
on the integral MACD with scaling variables 1, 1.2, 1.5, 2 and 3 applied on all
stocks of the current $S\&P100$ and $Eurostoxx50$ in the period from January
$1989$ until January $2016$. 

\section{Retracements}\label{sec:2}
\subsection{Distribution of the Retracement}
\begin{figure}[h]
  \centering
	\subfigure{\includegraphics[keepaspectratio=true,width=0.49\linewidth]{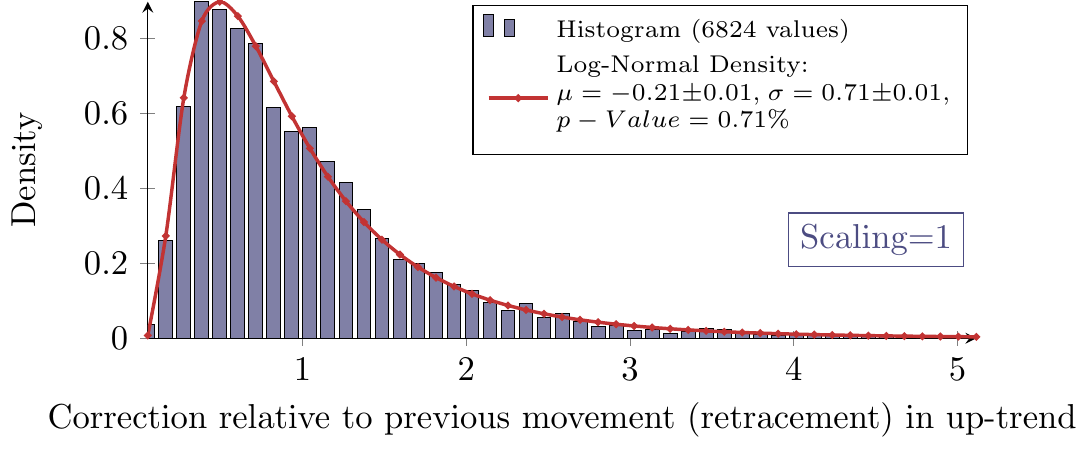}}
	\subfigure{\includegraphics[keepaspectratio=true,width=0.49\linewidth]{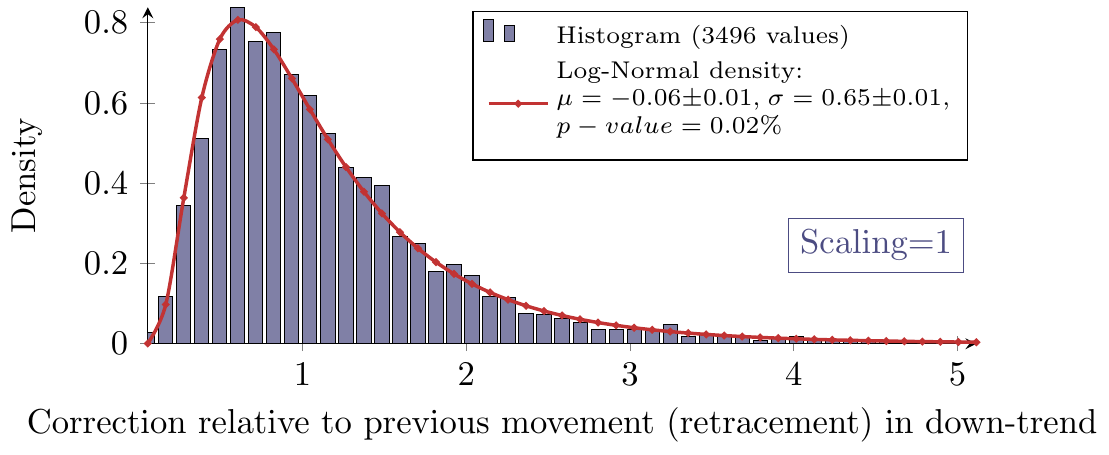}}
	\caption{Measured and log-normal-fit density of the retracement $X$ in an
	up-trend and down-trend with scaling $1$ for $S\&P100$ stocks. Each
	data set is visualized with a histogram from $0$ to $5$ with a bin size of $0.11$.}
\label{fig:histo_retr}
\end{figure}

\begin{figure}[h]
  \centering
	\subfigure{\includegraphics[keepaspectratio=true,width=0.49\linewidth]{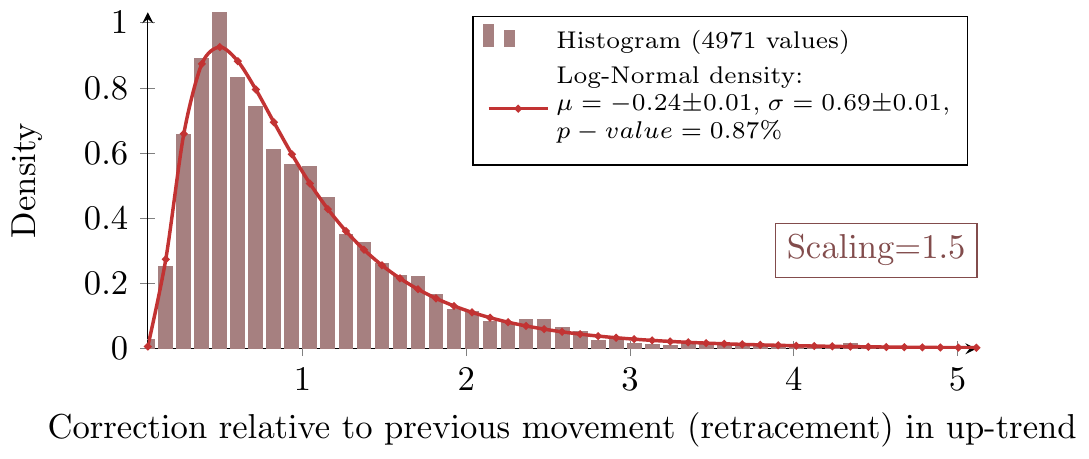}}
	\subfigure{\includegraphics[keepaspectratio=true,width=0.49\linewidth]{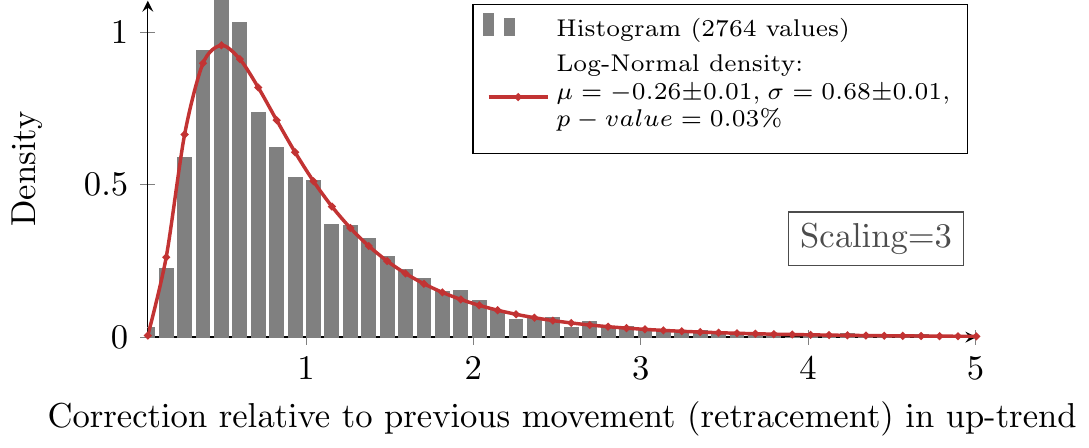}}
	\caption{Measured and log-normal-fit density of the retracement $X$ in an
	up-trend with scaling $1.5$ and $3$ for $S\&P100$ stocks. Each data set is visualized
with a histogram from $0$ to $5$ with a bin size of $0.11$.}
\label{fig:histo_retr_1}
\end{figure}

For all combinations of the regarded scalings and markets, the
measured retracement data shows the same characteristic distribution as seen in
Figures \ref{fig:histo_retr} and \ref{fig:histo_retr_1}. Indeed, they show the
typical asymmetric characteristic of a log-normal distribution which density is given by:
\begin{equation*}
f(x;\mu,\sigma)=\frac{1}{\sqrt{2\pi}\sigma
x}\exp{\left (-\frac{(\ln{(x)}-\mu)^2}{2\sigma^2}\right) },\quad x>0
\end{equation*}
for the retracement $X$ and with the (true) parameters $\mu$ and
$\sigma$. It is well known how to calculate moments of log-normally distributed
random variables. In this particular context, the median of the distribution $X$
equals $e^\mu$ and the mean is given by
\begin{equation*}
\E(X)=e^{\mu+\frac{\sigma^2}{2}}.
\end{equation*}
\par
To evaluate this distribution assumption the
maximum-likelihood-estimators (MLE) denoted
by $(\hat{\mu},\hat{\sigma})$ for the log-normal distribution are computed:
\begin{equation*}
\hat{\mu}:=\frac{1}{n}\sum_{i=1}^n \ln{x_i},\quad
\hat{\sigma}^2:=\frac{1}{n}\sum_{i=1}^n\left(\ln{(x_i)}-\mu\right)^2
\end{equation*}
with $x_i$ denoting the $n$ measured retracements. Furthermore,  the p-value
calculated with the Anderson-Darling test (recommended EDF test by Stephens in
\cite{Stephens}, chapter ``Test based on EDF statistics'') being applied to the
logarithmic transformed data is checked.
The such obtained values are summarized in Table \ref{table:retr}.\par
\begin{table}[h]
\centering
  \subfigure[$S\&P100$
  data]{\includegraphics[keepaspectratio=true,width=0.45\linewidth]{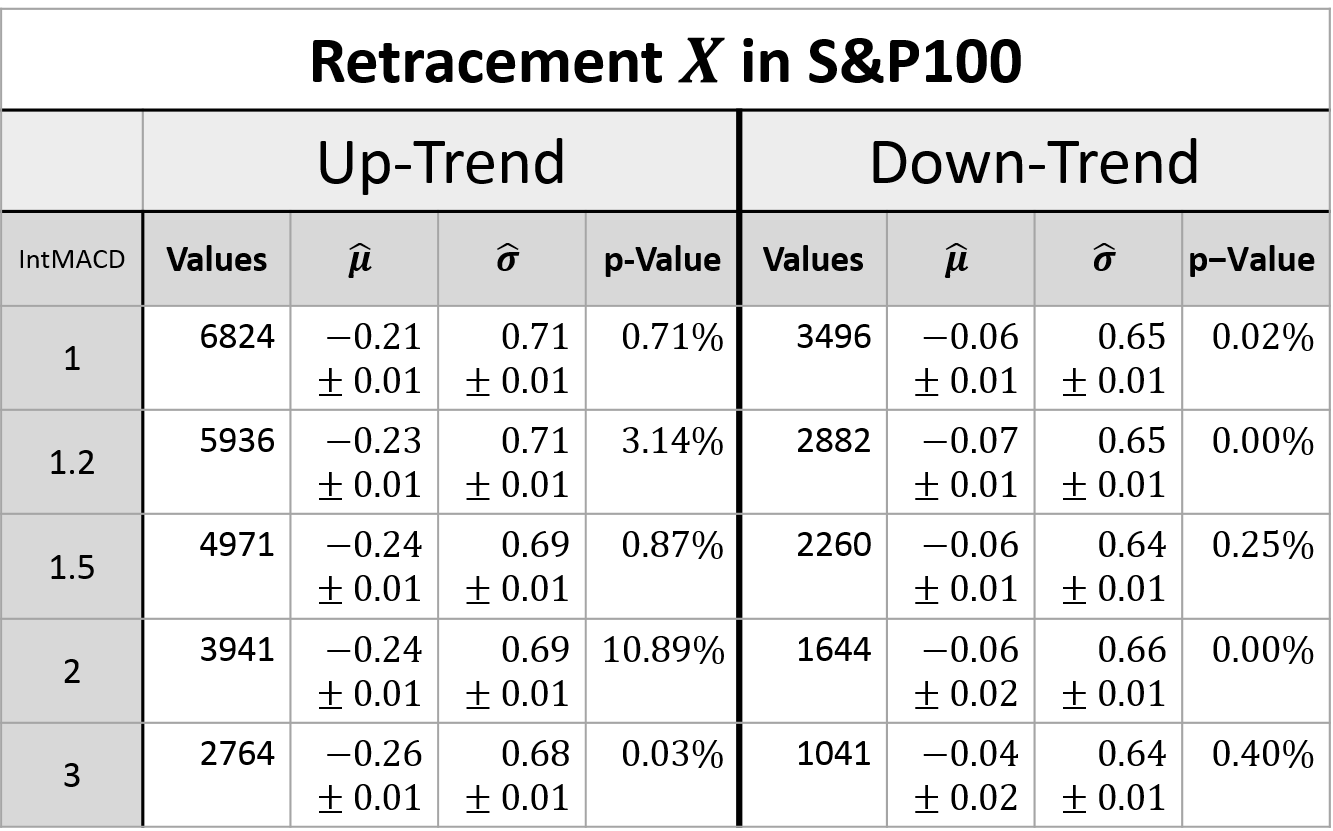}
  }
  \subfigure[$Eurostoxx50$ data]
  {\includegraphics[keepaspectratio=true,width=0.45\linewidth]{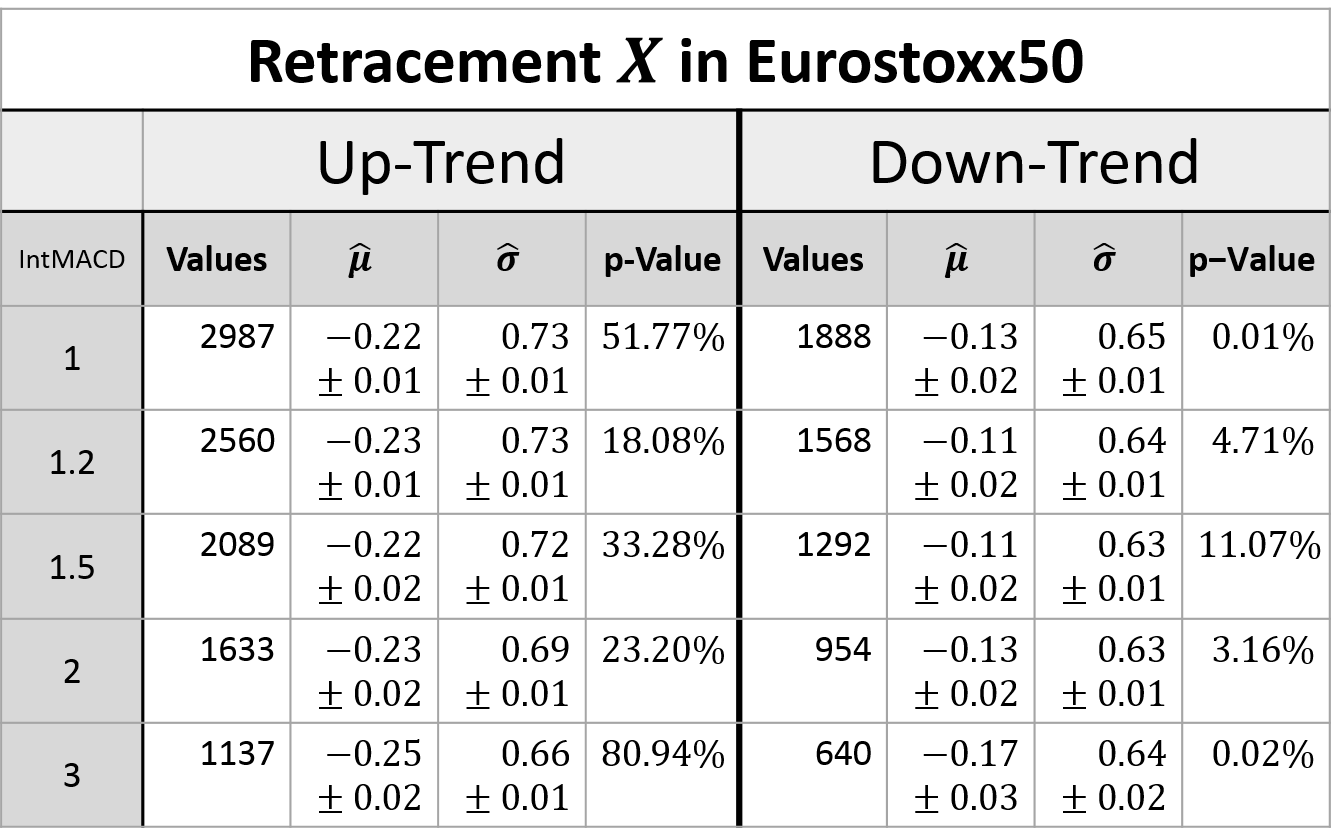}
  } \caption{Parameters of the log-normal fit for the retracement $X$ in up-
and down-trends.}
\label{table:retr}
\end{table}
The inconsistent p-values reveal that the log-normal model does not fit
perfectly to the measured retracement data. In fact, all histograms show a
slightly sharper density for the measured data with different intensities.
Consequently, for higher retracement values the log-normal model predicts
slightly less values than actually observed.
Besides this small systematical aberrations the log-normal model maps the
measurement very well -- especially for the $Eurostoxx50$. On top of that, the
log-normal model obviously fits much better to the retracement distribution than
to for instance daily returns of stock prices (see Fama \cite{FAMA63}).\par
To conclude the evaluation of the retracement alone, a fundamental
observation about the retracement can be made (based on the
log-normal assumption and the fit values derived from the data).
\begin{Observation}[Log-Normal model for the retracement]\label{beob_retr}
The parameters $\mu$ and $\sigma$ of the log-normal
distribution are more or less scale invariant for the retracement. In case of an
up-trend, the parameters are also market invariant.\par
Furthermore, the parameter $\mu$ is affected by the trend direction. It is
larger for down-trends, i.e.
the retracements in down-trends are overall more likely to be larger as in
up-trends. In spite of that, the parameter $\sigma$ is more or less invariant of
the trend direction.
\end{Observation}

\subsection{Delay after a Retracement}\label{sec:delay}
As already mentioned, the delay of the MinMax-process is
inevitable. Therefore, it will be evaluated in the same way as the
retracement. In order to be able to compare the delay $d_{abs}$ after a
retracement is recognized (as defined in (\ref{def_delay})) with the retracement
$X$ itself, both must have the same unit. So, the delay will also be considered in units of
the last movement. It will be denoted as random variable $D_X$:
\begin{equation*}
D=D_X=\frac{d_{abs}}{Movement}.
\end{equation*}
It should be noted that (at this point) there is no statement made on whether or
not $D_X$ may somehow depend on the preceding retracement $X$. The
notation with subscript $X$ is only used to denote the delay after a
retracement and to distinguish it from other delays to come.\par
\begin{figure}[h]
  \centering
	\includegraphics[keepaspectratio=true,width=0.7\linewidth]{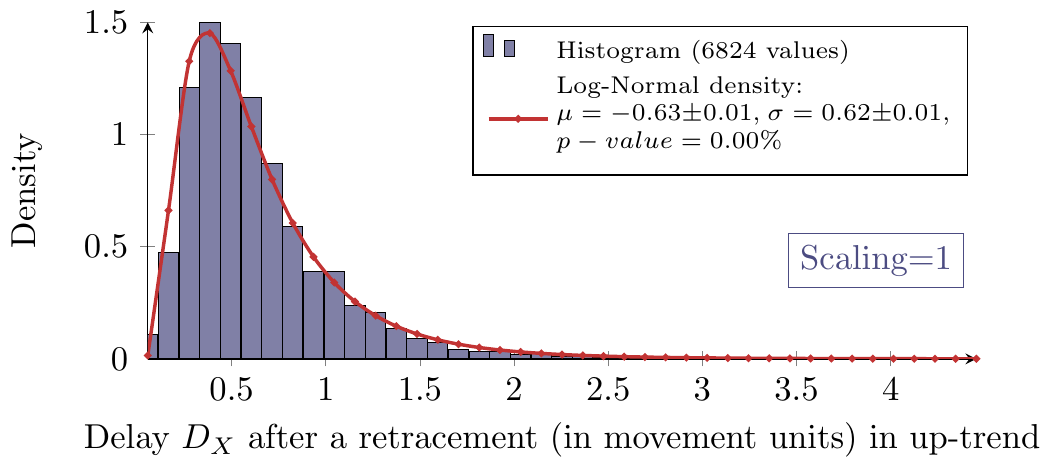}
	\caption{Measured and log-normal-fit density of the delay in an up-trend
with scaling $1$ for $S\&P100$ stocks. The data is visualized with a histogram
from $0$ to $5$ with a bin size of $0.11$.}
\label{fig:delay}
\end{figure}

Again, the
measured delay data shows the characteristic of a log-normal distribution for
each combination of scaling and market as exemplarily shown in Figure
\ref{fig:delay}.
However -- as expressed by the significant p-values in Table \ref{table:delay}
-- the log-normal assumption is definitively wrong.
\begin{table}[h]
\centering
  \subfigure[$S\&P100$
  data]{\includegraphics[keepaspectratio=true,width=0.45\linewidth]{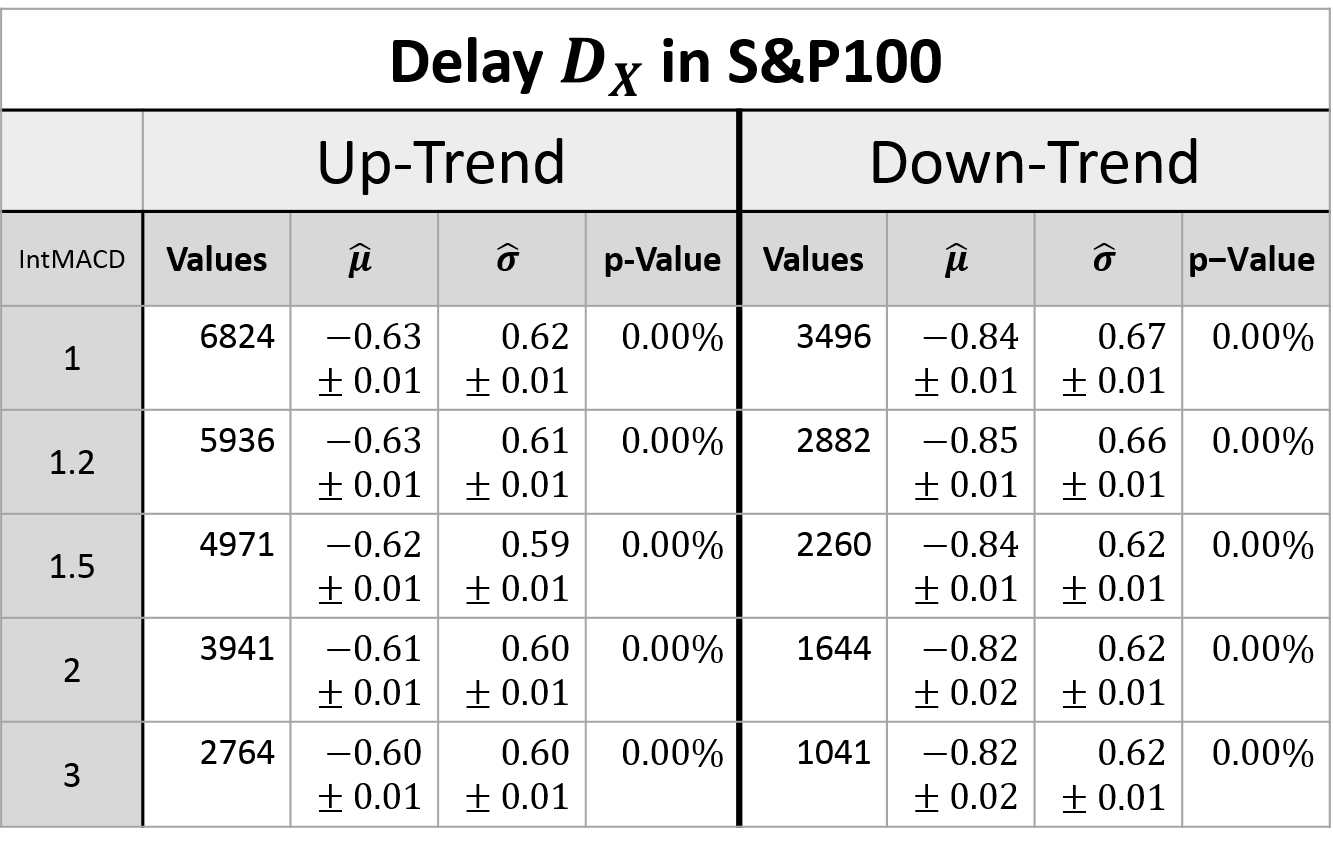}
  } \subfigure[$Eurostoxx50$
  data]{\includegraphics[keepaspectratio=true,width=0.45\linewidth]{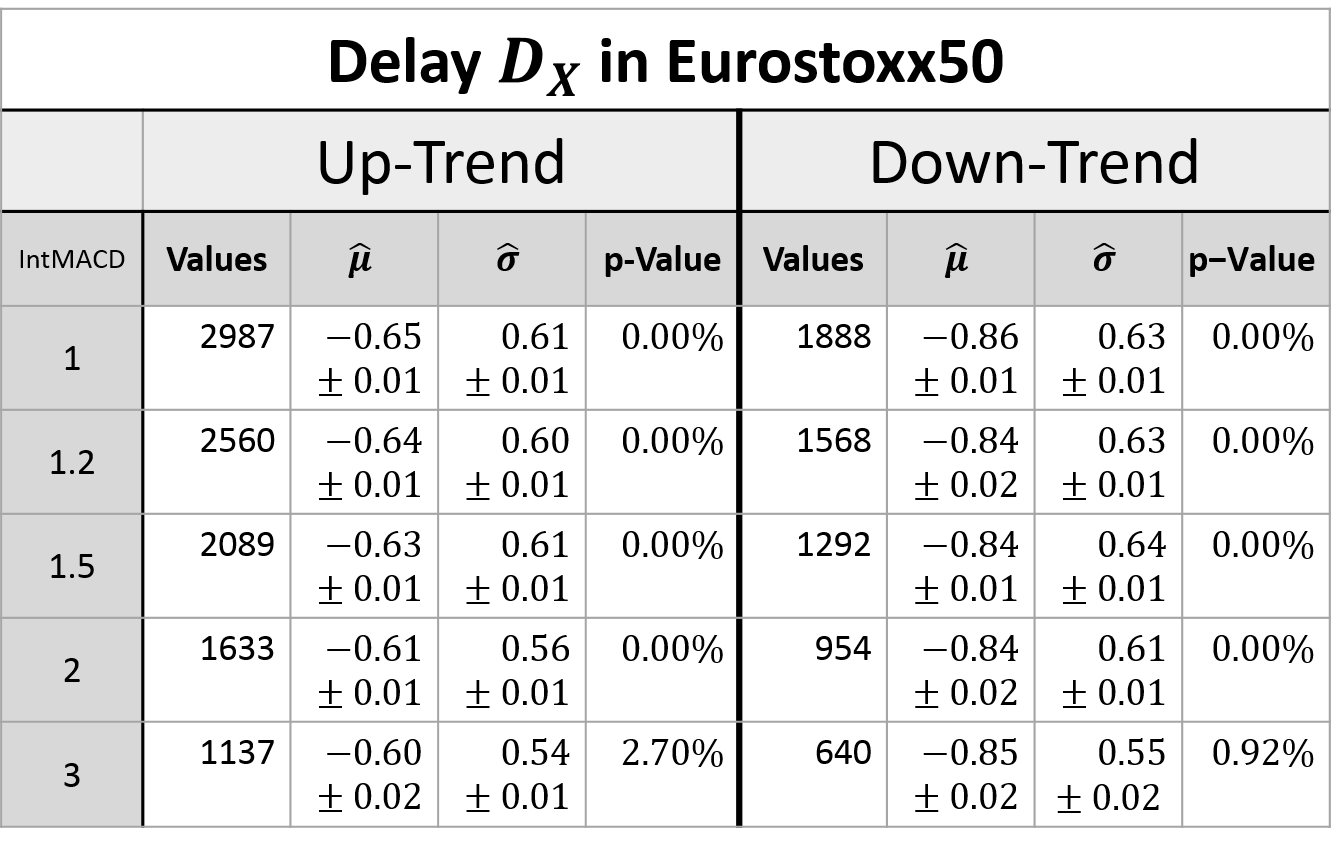} } 
  \caption{Parameters of the log-normal fit for the delay $D_X$ in
  up- and down-trends.}
\label{table:delay}
\end{table}
The histograms show a systematical
deviation in regard to skewness. The measured delays have a less positive
skewness than predicted by the model.\par
Besides this systematical aberration the log-normal model maps the
characteristic of the measured delay well enough such that it will be used for
the following analysis.\par

The retracement and the delay can be considered as one sequence. We therefore
look for a combined log-normal distribution of retracement and delay. In this
context it is important to evaluate the estimator of the correlation $\rho$
between the logarithm of the two variables, i.e.
\begin{equation*}
\hat{\rho}_{\ln{X},\ln{D}}=\frac{\frac{1}{n}\sum_{i=1}^n
(\ln({x_i})-\hat{\mu}_X)(\ln{(d_i)}-\hat{\mu}_D)}{\hat{\sigma}_X\cdot
\hat{\sigma}_D}
\end{equation*}
for measured values $(x_i,d_i)$. The estimated values of $\hat{\rho}$ are given
in Table \ref{table:cor}.
\begin{table}[h]
\centering
  \subfigure[$S\&P100$
  data]{\includegraphics[keepaspectratio=true,width=0.35\linewidth]{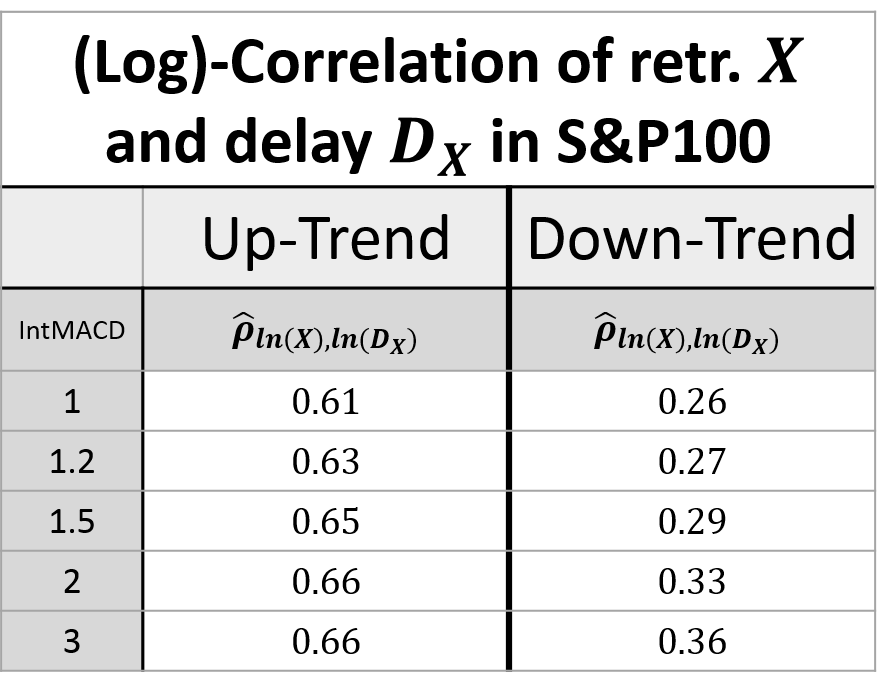}
  } \subfigure[$Eurostoxx50$
  data]{\includegraphics[keepaspectratio=true,width=0.35\linewidth]{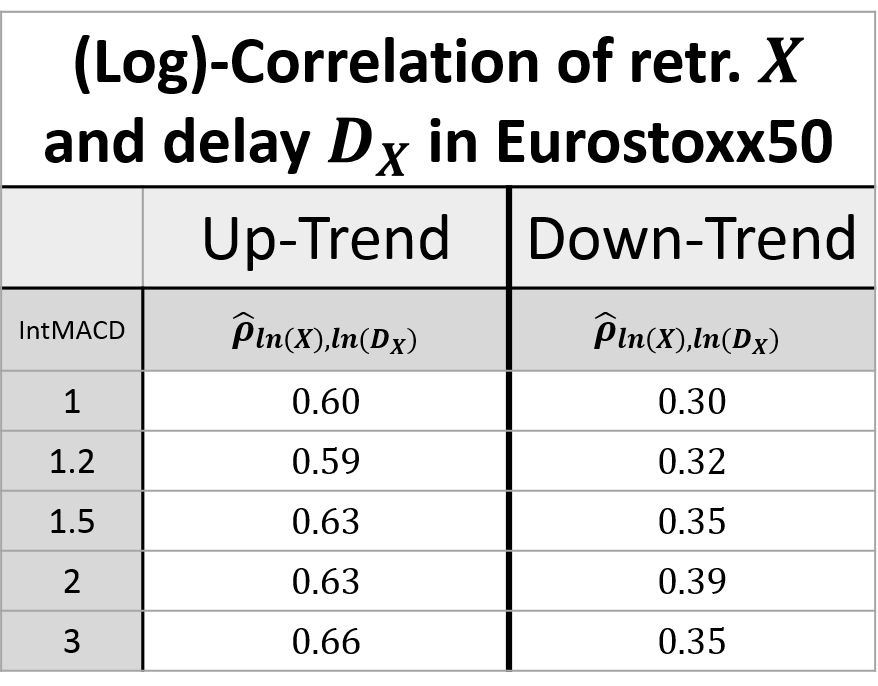}
  } \caption{Correlation between the logarithms of the retracement and
  delay (after retr.) in up- and down-trends.}
\label{table:cor}
\end{table}
It shows, that the retracement and the following delay are indeed positively
correlated (regarded in the same units). This way it is possible to give a
joint bivariate log-normal distribution for the retracement $X$ and the
delay $D$ (both in units of the preceding movement) by virtue of its density
function.
\begin{eqnarray}\label{eqn_xd}
&\,&f_{X,D}\left(x,d;\mu_X,\mu_{D},\sigma_X,\sigma_{D},\rho\right)=\frac{1}{2\pi
x d \sigma_X \sigma_{D} \sqrt{1-\rho^2}}*\\
&*&\exp\left[-\frac{1}{2(1-\rho^2)}\left(\frac{(ln(x)-\mu_X)^2}{\sigma_X^2}+\frac{(ln(d)-\mu_{D})^2}{\sigma_{D}^2}-2\rho\frac{(ln(x)-\mu_X)(ln(d)-\mu_{D})}{\sigma_X\sigma_{D}}\right)\right].
\nonumber
\end{eqnarray}
For calculations based on this distribution, the (true) parameters
$\mu_X,\mu_{D},\sigma_X,\sigma_{D}$ and $\rho$ must be replaced by their
estimators.\par
Finally, the concluding observation regarding the retracement can
be expanded by the delay part.
\begin{Observation}[Log-Normal model for the retracement and
delay]\label{beob_retr_delay} The parameters $\mu$ and $\sigma$ of the
log-normal distribution are more or less scale invariant for the retracement and
the delay.
In case of an up-trend, the parameters are also market invariant.\par
Furthermore, the parameter $\mu$ is affected by the trend direction. In case
of the retracement it is larger for down-trends whereas it is
smaller for down-trends in case of the delay. In spite of that, the parameter
$\sigma$ is more or less invariant of the trend direction.\par 
Finally, the correlation between the logarithms of the retracement and the
delay are close to scale and market invariant while the correlation in up-trends is significantly larger than in down-trends.
\end{Observation}
\subsection{Fibonacci Retracements}
\begin{figure}[h]
  \centering
	\includegraphics[keepaspectratio=true,width=0.7\linewidth]{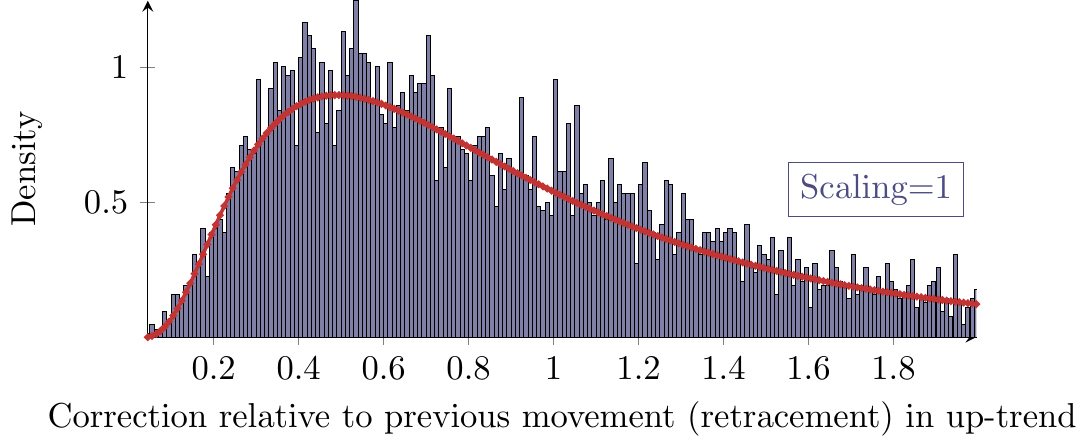}
	\caption{More detailed (finer) histogram of Figure \ref{fig:histo_retr} with
scaling $1$ from $0$ to $2$ with a bin size of $0.01$.}
\label{fig:fibo}
\end{figure}

A propagated idea in the field of technical analysis for dealing with
retracements is the concept of so called \textit{Fibonacci Retracements}.
Based on specific retracement levels derived from several powers of the
inverse of the golden ratio one wants to make a priori predictions for future
retracement values. Obviously, this assumes that there are such significant
retracement values. However, the evaluation of the retracement above reveals
that there are no levels with a great statistical significance but the
retracements follow a continuous distribution overall. Even a finer histogram
as shown in Figure \ref{fig:fibo} does not reveal any significant
retracements.\par
On the assumption that there are specific values with
statistical significance in some regard, then the $100\%$-level would be most
significant.
For a closer look on significant retracement levels see \cite{IFTA_FIBO}.

\subsection{Duration of the Retracement}
\begin{figure}[h]
  \centering
	\subfigure[Retracement duration $Y$ in up-trend ($Mean=13.0$)]
	{\includegraphics[keepaspectratio=true,width=0.45\linewidth]{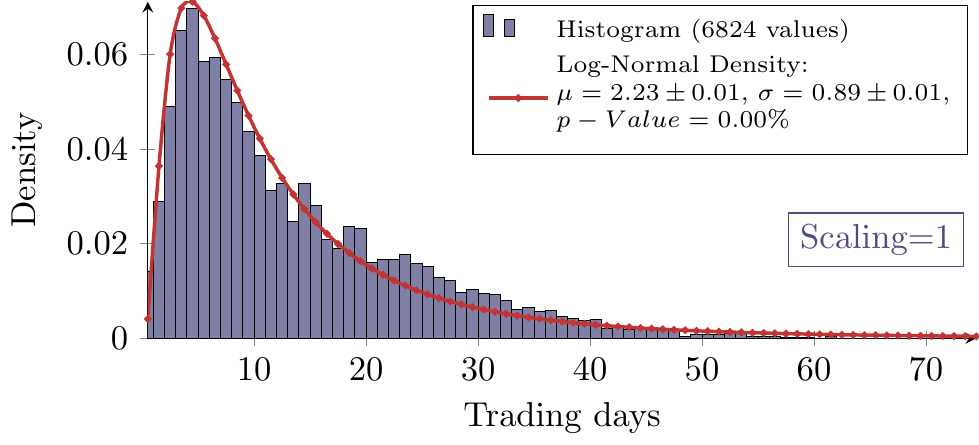}
	}
	\subfigure[Retracement duration $Y$ in down-trend ($Mean=17.6$)]
	{\includegraphics[keepaspectratio=true,width=0.45\linewidth]{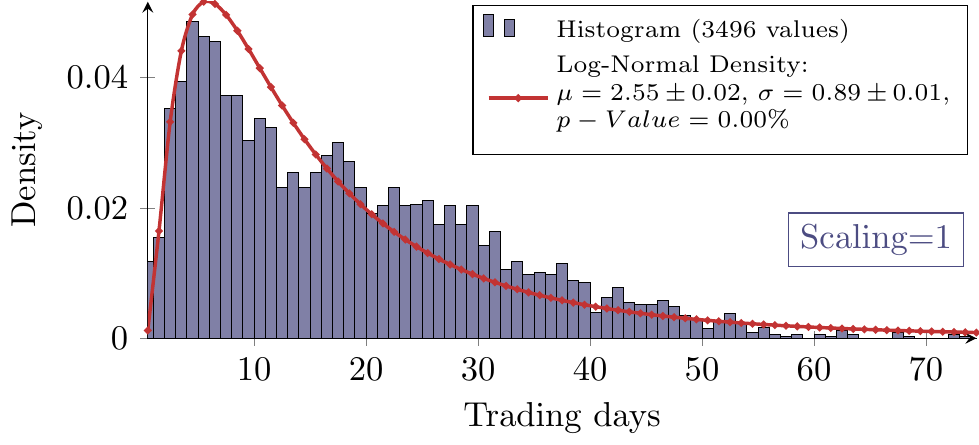}
	} \caption{Measured and log-normal-fit density of the retracement duration
	in up- and down-trends (left and right resp.) with scaling $1$
	for $S\&P100$ stocks. The data is visualized with a histogram with a bin size
	of $1$.}
\label{fig:histo_duration}
\end{figure}

Beside the retracement, the \textit{duration of a trend correction}
denoted by $Y$ is also of interest. It is given by the difference in trading
days between the last $P2$ and the new $P3$ (see Figure \ref{fig:trend}.(b)).
The distributions of the retracement duration overall show the asymmetric
log-normally-like behavior as exemplarily shown in Figure \ref{fig:histo_duration}. However, the goodness of the log-normal assumption is obviously worse as in the case of the retracement itself. In particular, the measured densities of the retracement duration in a down-trend all show significant aberrations from the log-normal model.\par
\begin{figure}[p]
  \centering
	\includegraphics[keepaspectratio=true,width=0.75\linewidth]{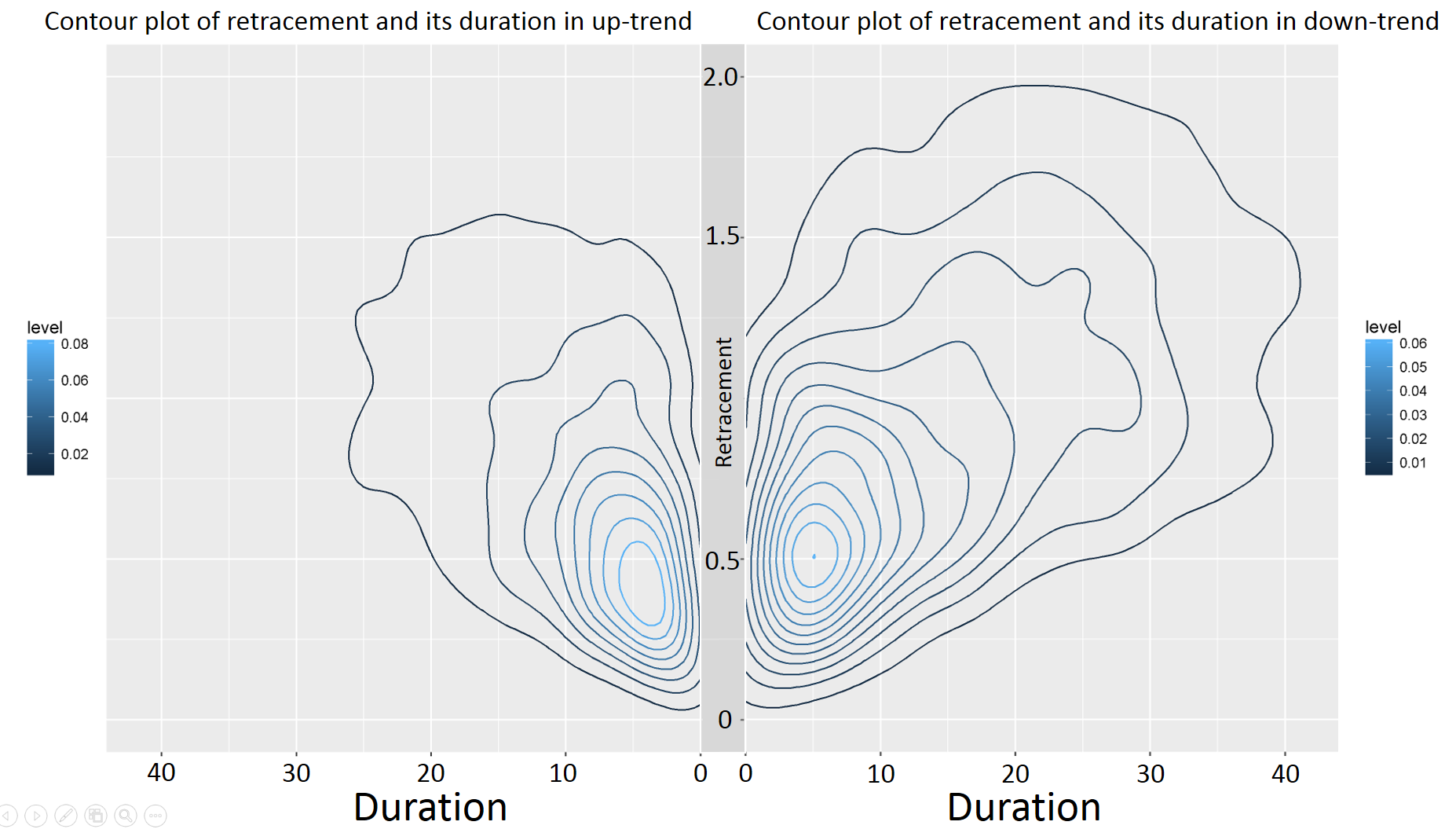}
	\caption{Contour plot of the joint density of the retracement value and its
duration in up- and down-trends (left and right resp.) with scaling $1$ for $S\&P100$ stocks.}
\label{fig:butterfly}
\end{figure}

\begin{table}[h]
\centering
  \subfigure[$S\&P100$
  data]{\includegraphics[keepaspectratio=true,width=0.35\linewidth]{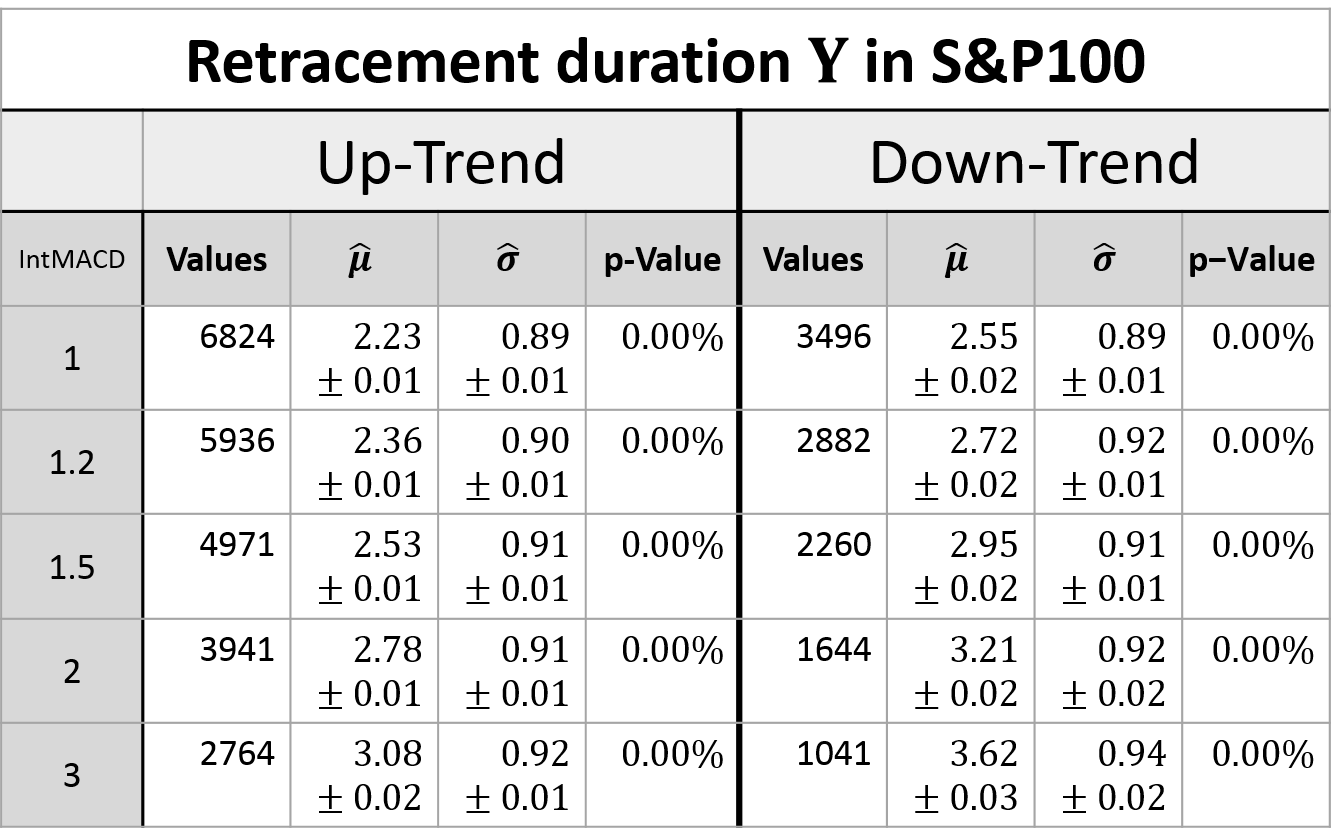}
  } \subfigure[$Eurostoxx50$
  data]{\includegraphics[keepaspectratio=true,width=0.35\linewidth]{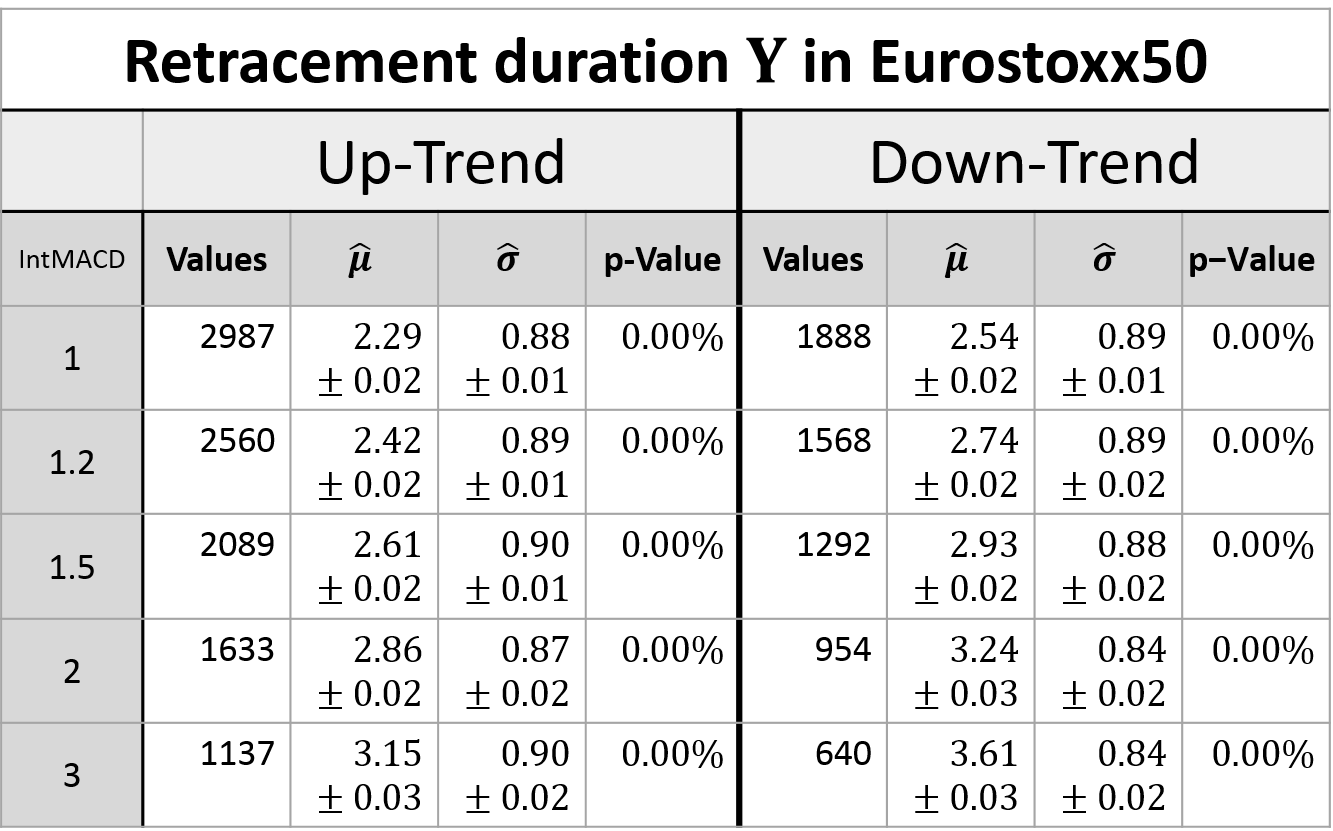}
  } \caption{Parameters of the log-normal fit for the retracement duration in
  up- and down-trends.}
\label{table:retr_dur}
\end{table}
Since every retracement value is associated with a duration, the joint
distribution of the retracement and its duration can be examined (see Tables
\ref{table:retr_dur} and \ref{table:retr_dur_cor}).
\begin{table}[p]
\centering
  \subfigure[$S\&P100$
  data]{\includegraphics[keepaspectratio=true,width=0.3\linewidth]{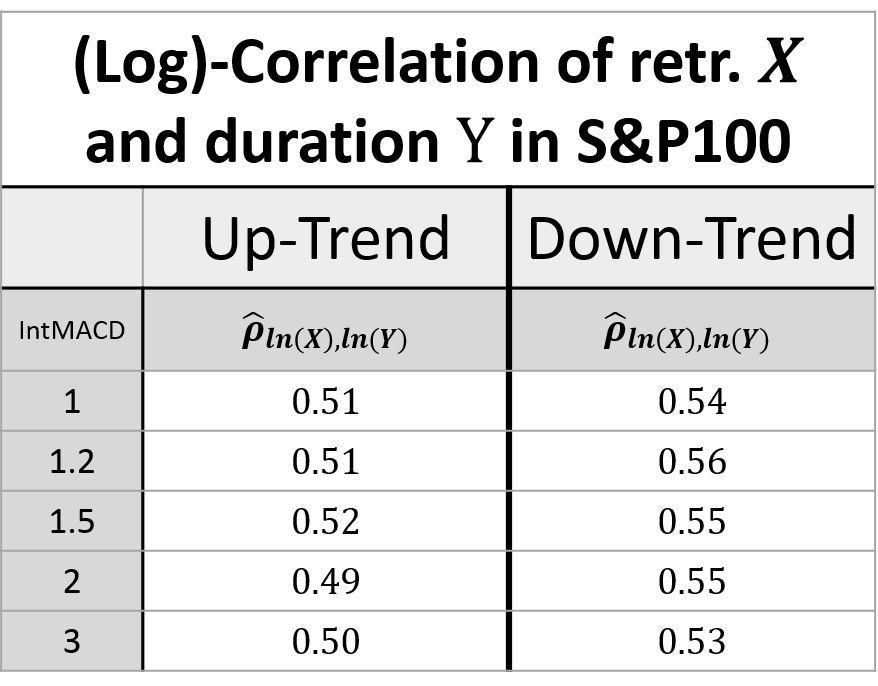}
  } \subfigure[$Eurostoxx50$
  data]{\includegraphics[keepaspectratio=true,width=0.3\linewidth]{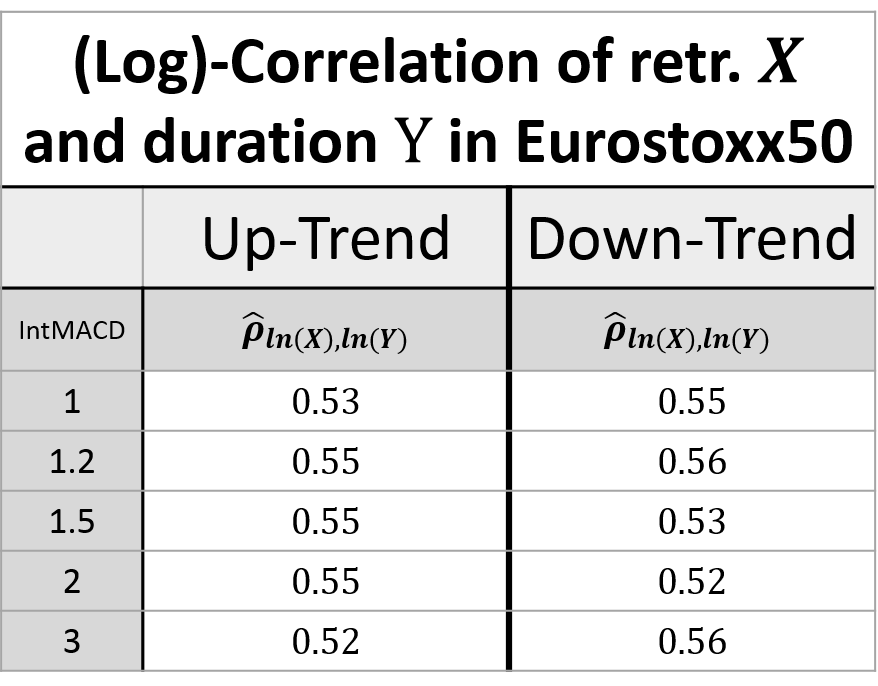}
  } \caption{Correlation between the logarithms of the retracement and its
  duration in up- and down-trends.}
\label{table:retr_dur_cor}
\end{table}
Figure \ref{fig:butterfly} exemplarily shows that the retracement in down-trends tends
to have higher values as in up-trends. This was already seen in Table
\ref{table:retr} and Observation \ref{beob_retr}. However, Figure \ref{fig:butterfly} exemplarily
also shows that the retracement in down-trends has larger durations
compared to the retracement in up-trends.
\newpage
\section{Movement and Correction}\label{sec:3}
\subsection{Distribution of Relative Movements and Corrections}
\begin{figure}[p]
  \centering
	\subfigure[relative
	Movement]{\includegraphics[keepaspectratio=true,width=0.45\linewidth]{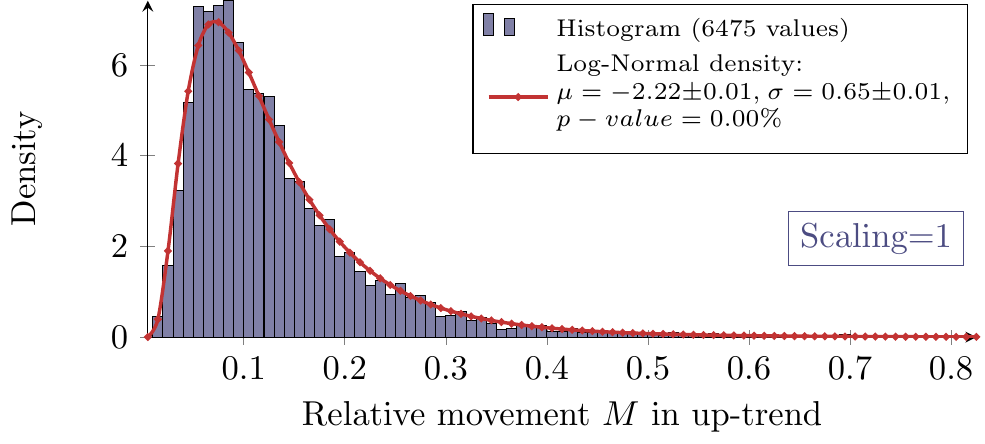}
	} \subfigure[relative
	Correction]{\includegraphics[keepaspectratio=true,width=0.45\linewidth]{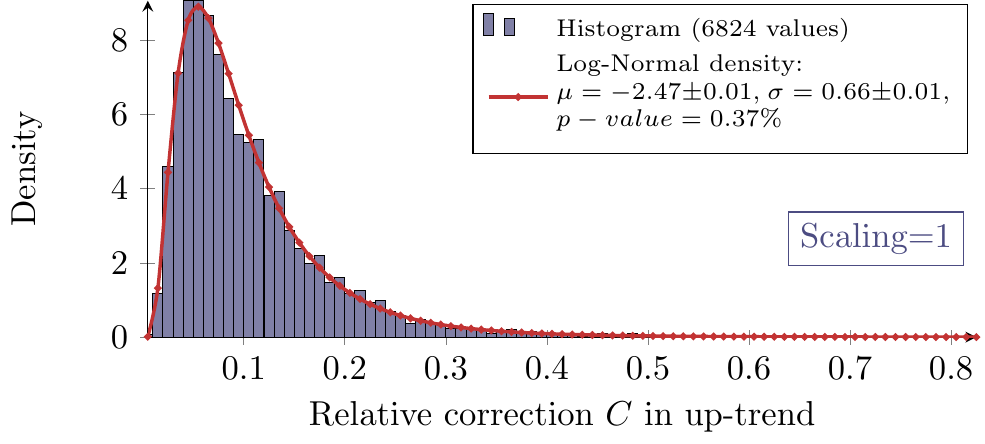}
	} \caption{Measured and log-normal-fit density of the relative movement (left) and relative correction (right) in an up-trend with scaling $1$ for $S\&P100$
stocks.
The data is visualized via histogram from $0$ to $1$ with a bin size of $0.01$.}
\label{fig:histo_bew}
\end{figure}

As before, all of the measurements show the same characteristic distribution
-- whether relative movement (\ref{eqn_bew}) or relative correction
(\ref{eqn_kor}).
As before, the histograms conclude the log-normal assumption (see Figure
\ref{fig:histo_bew}).
Again, the log-normal model does not match the measured data perfectly, but
often fails to map the sharp peaks and fat tails. This observation is confirmed
by the fluctuating p-values (see Table \ref{table:bew} and \ref{table:kor}).

\begin{table}[h]
\centering
  \subfigure[$S\&P100$
  data]{\includegraphics[keepaspectratio=true,width=0.4\linewidth]{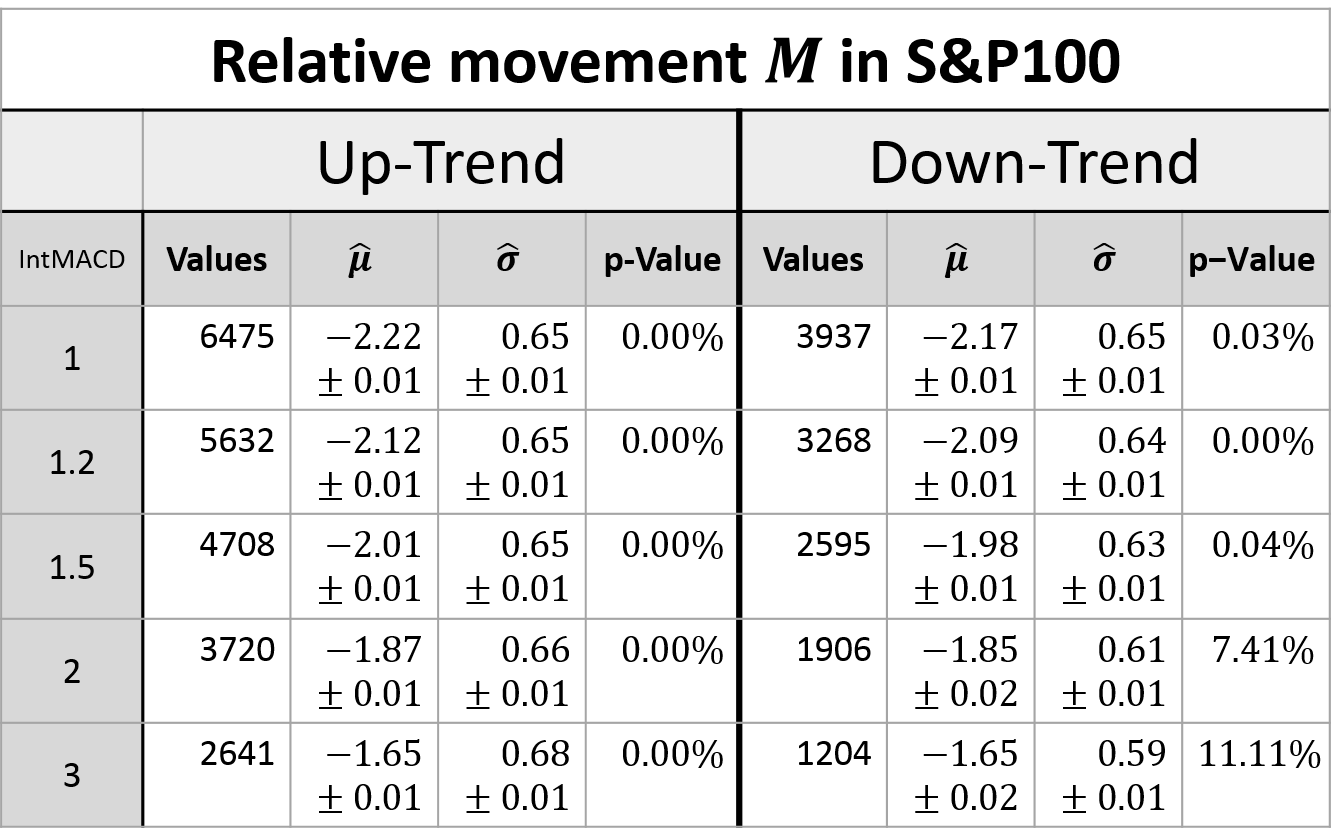}
  } \subfigure[$Eurostoxx50$
  data]{\includegraphics[keepaspectratio=true,width=0.4\linewidth]{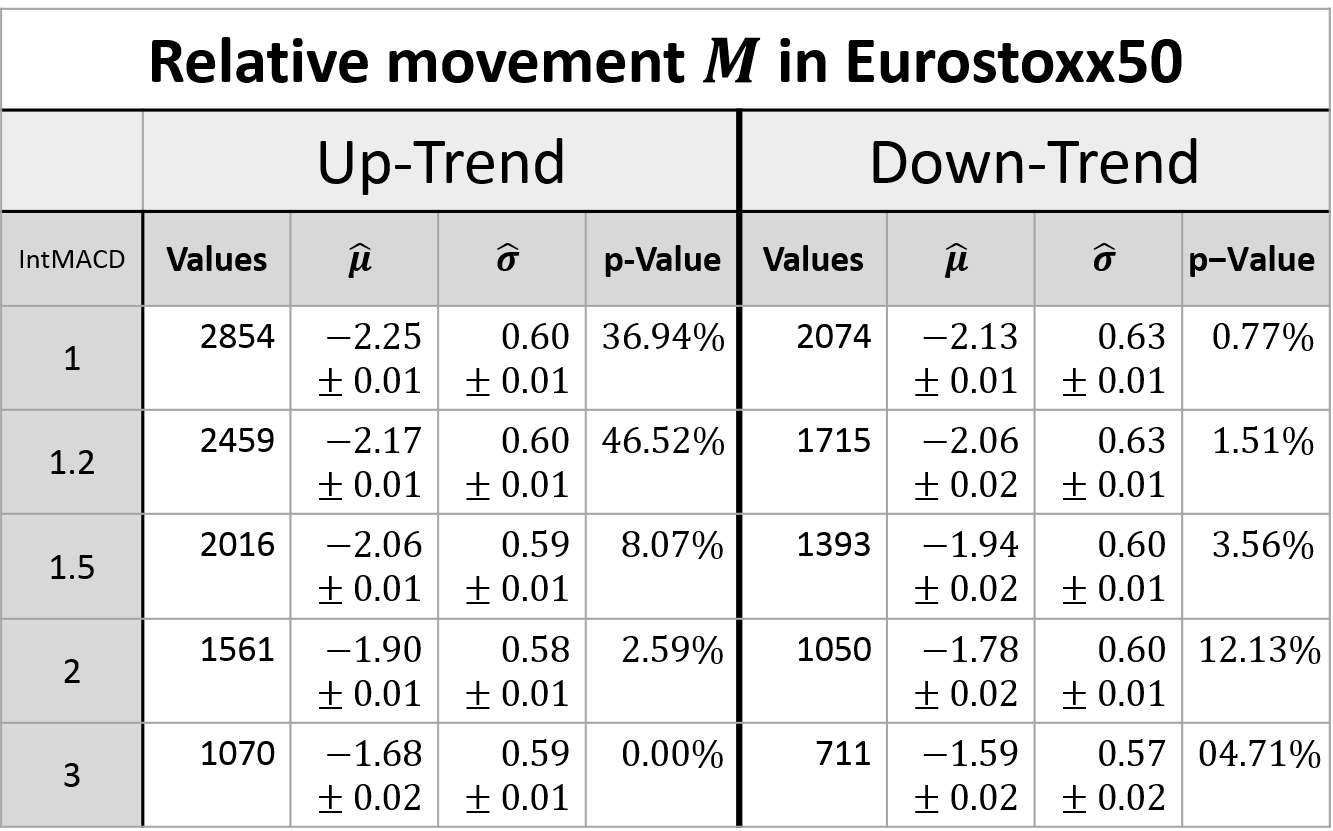}
  } \caption{Parameters of the log-normal fit for the relative movement in up- and down-trends.}
\label{table:bew}
\end{table}

\begin{table}[h]
\centering
  \subfigure[$S\&P100$
  data]{\includegraphics[keepaspectratio=true,width=0.4\linewidth]{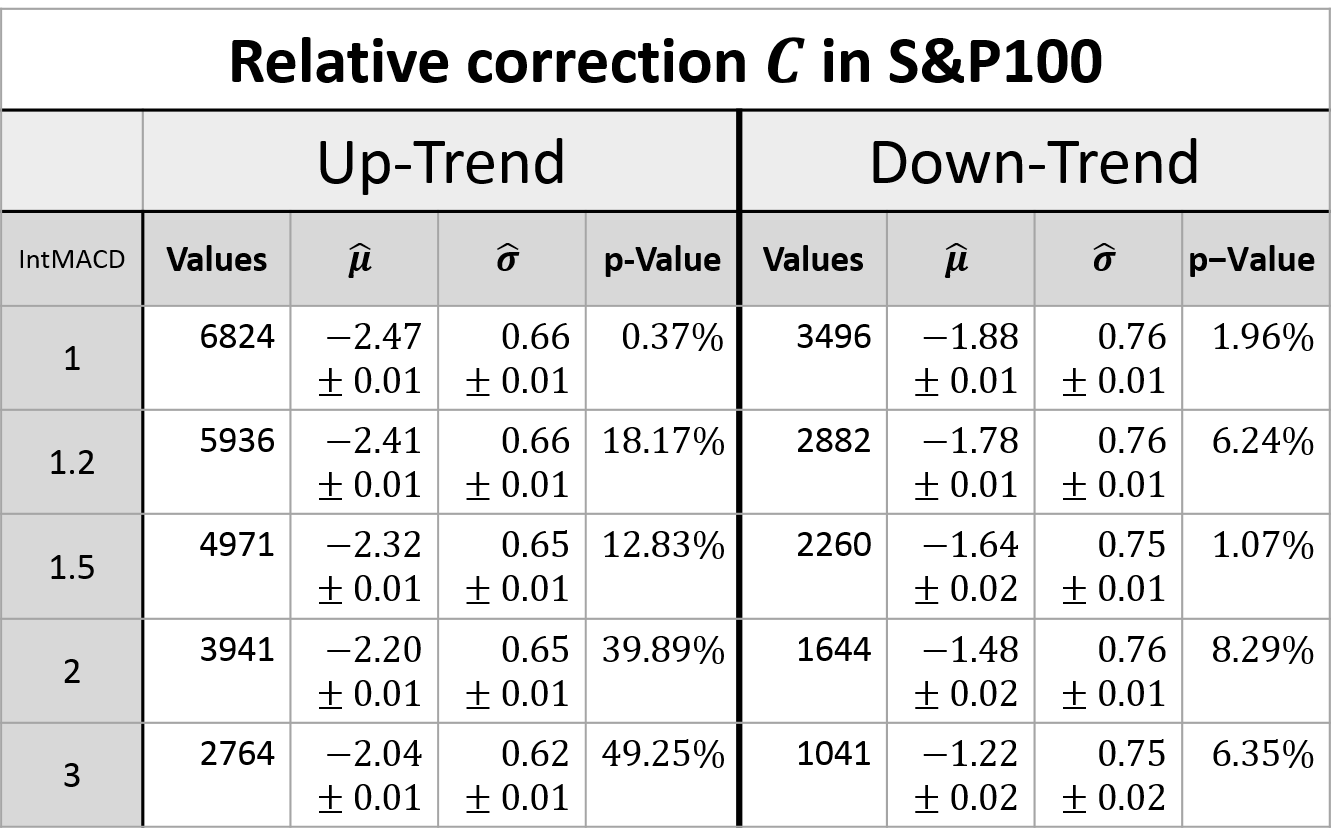}
  } \subfigure[$Eurostoxx50$
  data]{\includegraphics[keepaspectratio=true,width=0.4\linewidth]{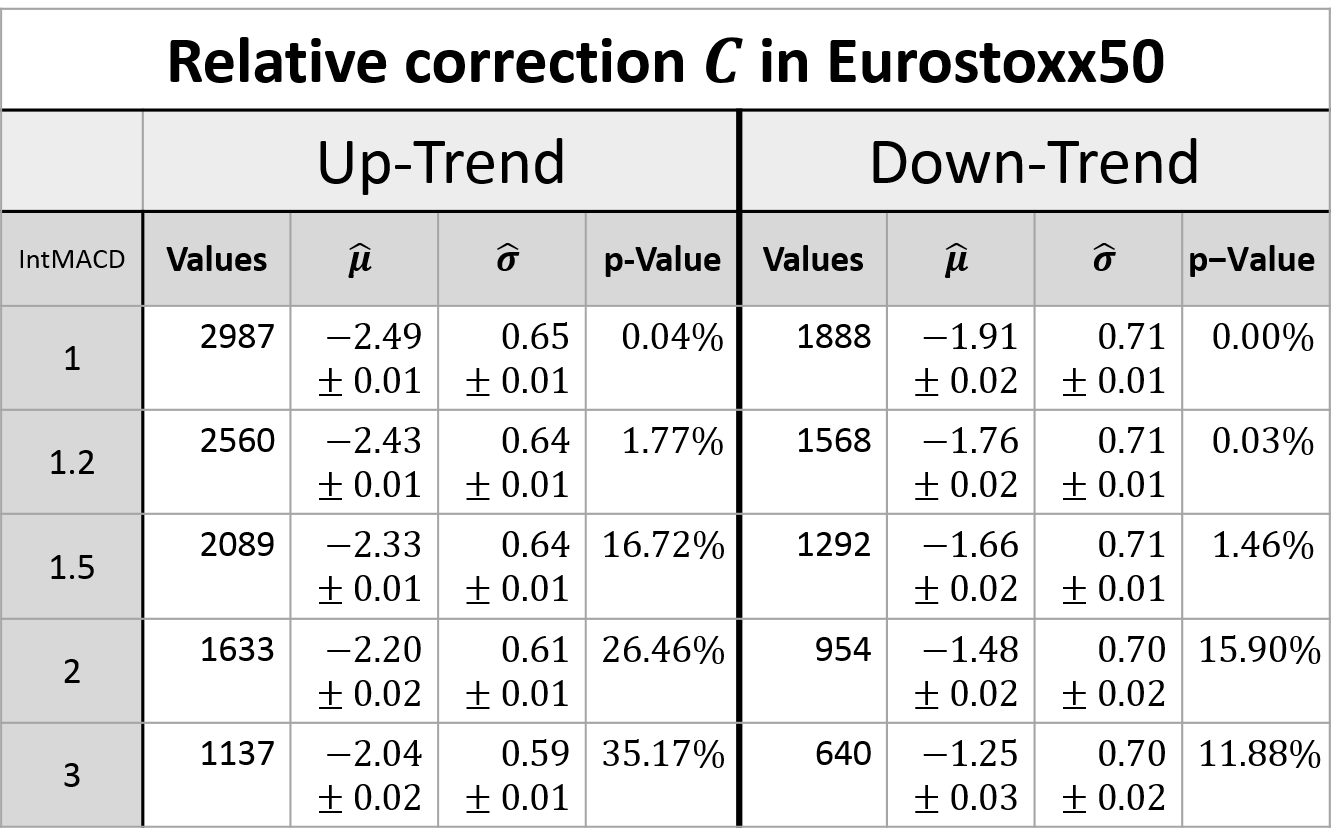}
  } \caption{Parameters of the log-normal fit for the relative correction in up- and down-trends.}
\label{table:kor}
\end{table}
Consequently, based on the results of Table  \ref{table:bew} and
\ref{table:kor} a fundamental observation can be made which differs from
the retracement's one.
\begin{Observation}[Log-Normal model for the rel.
movement/correction]\label{beob_bew}
The parameters of the log-normal
distribution $\mu$ and $\sigma$ are market invariant for the relative movement
and correction. Furthermore, the $\sigma$ parameter is more or less scale
invariant while $\mu$ increases for increasing scaling for the relative movement
and correction, i.e. the relative movement and correction are more likely to
be larger for higher scaling parameter.\par
The parameters $\mu$ and $\sigma$
are also more or less trend direction invariant for the relative movement. In case of a relative correction, however, the direction of a
trend affects these parameters: Both are larger in case of a down-trend. 
\end{Observation}
The dependency between the $\mu$ parameter and the scaling has already been expected as outlined above.
Obviously, higher scalings yield more significant movements and
corrections. Therefore, to reflect this, the $x$-position of the density peak
has to increase when the scaling increases. The dependency of the $\mu$ parameter on the trend direction has already been
observed for the retracement (see Observation \ref{beob_retr}).
\subsection{Delay after relative Movements and Corrections}
As before, the delay $d$ also has to be taken into account. Its absolute value
is given by
\begin{equation*}
d_{abs}=|(\text{new extremum})-(\text{Close when new extremum is subsequently
detected})|
\end{equation*}
as defined in (\ref{def_delay}).
Here, the
unit for the delay is the last extreme value. This means for up-trends, the delay after the
relative movement is given by
\begin{equation*}
D_M:=\frac{d_{abs}}{last Low}
\end{equation*}
while the delay after the relative correction is given by
\begin{equation*}
D_C:=\frac{d_{abs}}{last High}.
\end{equation*}
In both cases, it is sometimes abbreviated as the \textit{relative
delay} and has the same unit as the relative movement (\ref{eqn_bew}) and relative correction
(\ref{eqn_kor}), respectively.\par
\begin{figure}[h]
  \centering
	\subfigure[Delay $D_M$ after relative
	movement]{\includegraphics[keepaspectratio=true,width=0.45\linewidth]{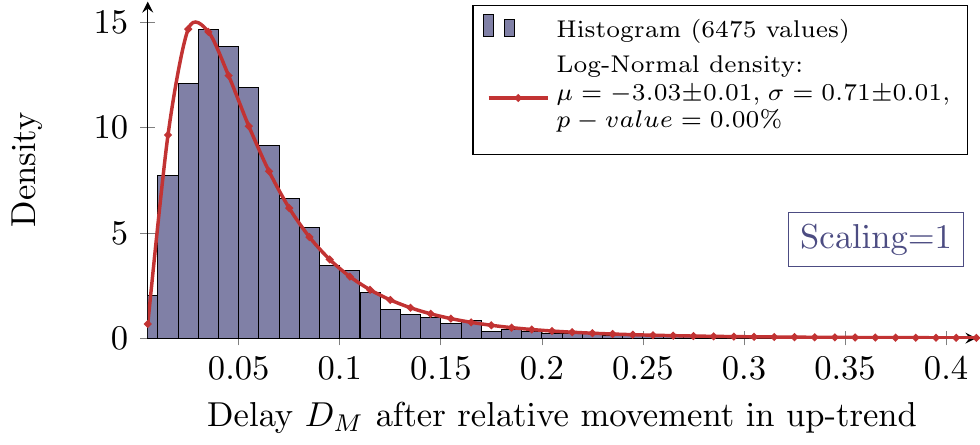}
	} \subfigure[Delay $D_C$ after relative
	correction]{\includegraphics[keepaspectratio=true,width=0.45\linewidth]{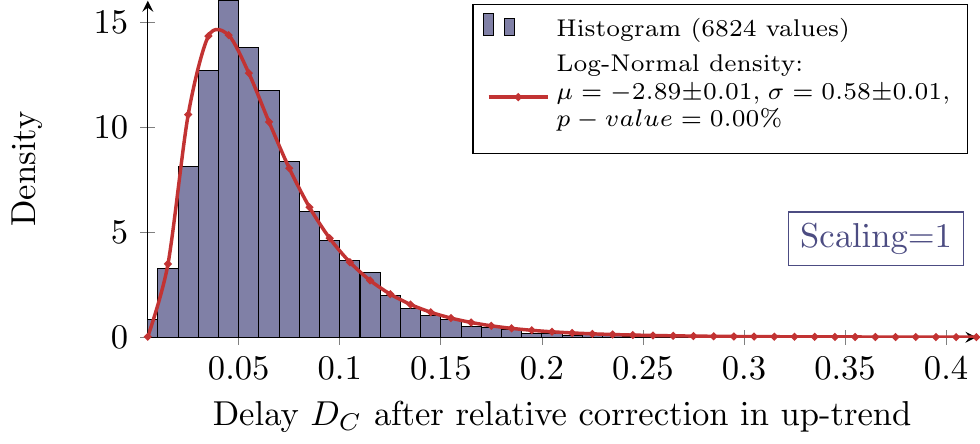}
	} \caption{Measured and log-normal-fit density of the relative delay after a relative movement $M$ (left) and after a relative correction $C$
	(right) in an up-trend with scaling $1$ for $S\&P100$ stocks.
The data is visualized with a histogram from $0$ to $1$ with a bin size of
$0.01$.}
\label{fig:bew_delay}
\end{figure}

Eventually, as shown in Figure
\ref{fig:bew_delay} the relative delay inherits the same characteristics as
known from the delay for the retracement.
As before, the model's skewness is slightly too positive. Consequently, the
conclusion is also the same. The model matches the measurements well enough to
be the base for further analysis.\par
\begin{table}[p]
\centering
  \subfigure[$S\&P100$
  data]{\includegraphics[keepaspectratio=true,width=0.45\linewidth]{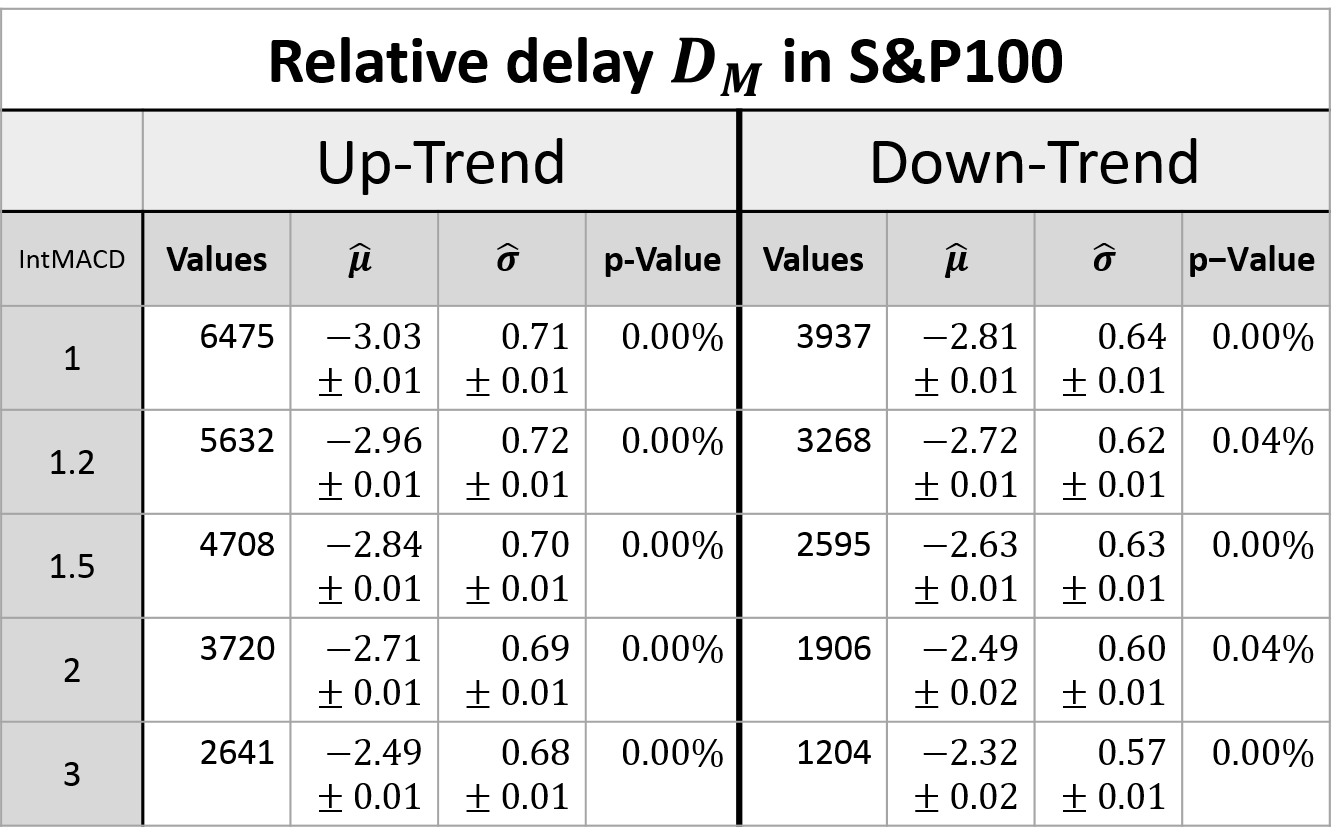}
  }
  \subfigure[$Eurostoxx50$
  data]{\includegraphics[keepaspectratio=true,width=0.45\linewidth]{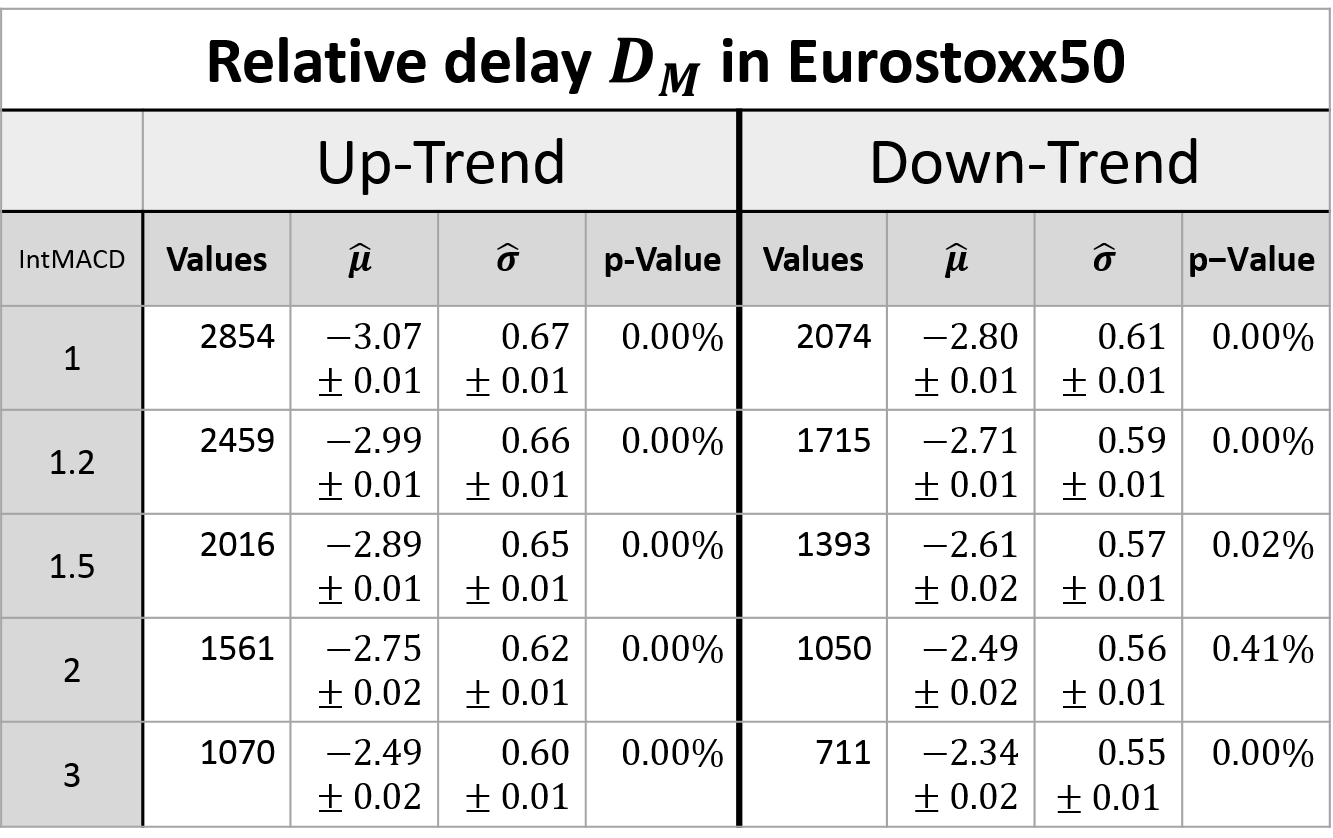}
  } \caption{Parameters of the log-normal fit for the relative delay $D_M$
  after a movement in up- and down-trends.}
\label{table:bew_delay}
\end{table}

\begin{table}[p]
\centering
  \subfigure[$S\&P100$
  data]{\includegraphics[keepaspectratio=true,width=0.45\linewidth]{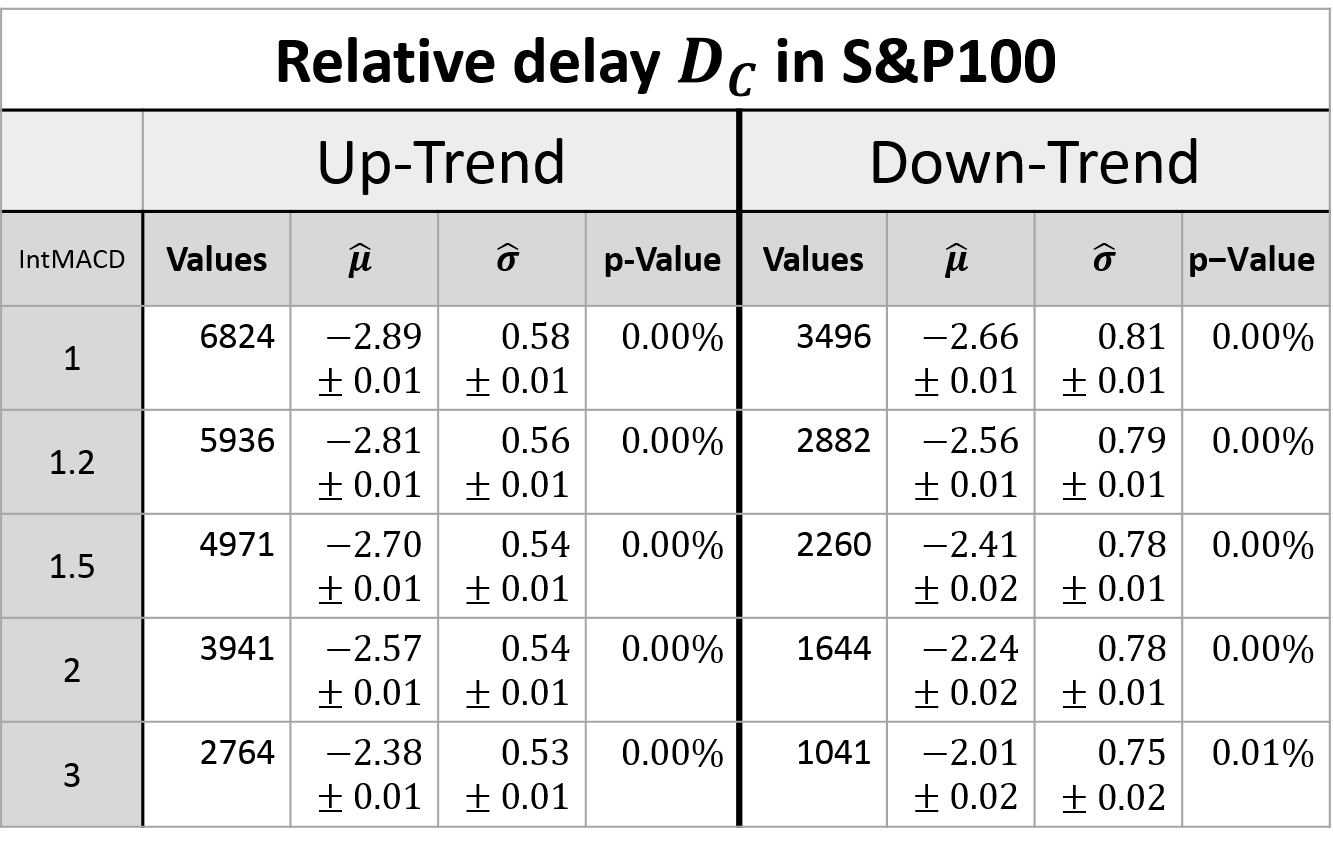}
  }
  \subfigure[$Eurostoxx50$
  data]{\includegraphics[keepaspectratio=true,width=0.45\linewidth]{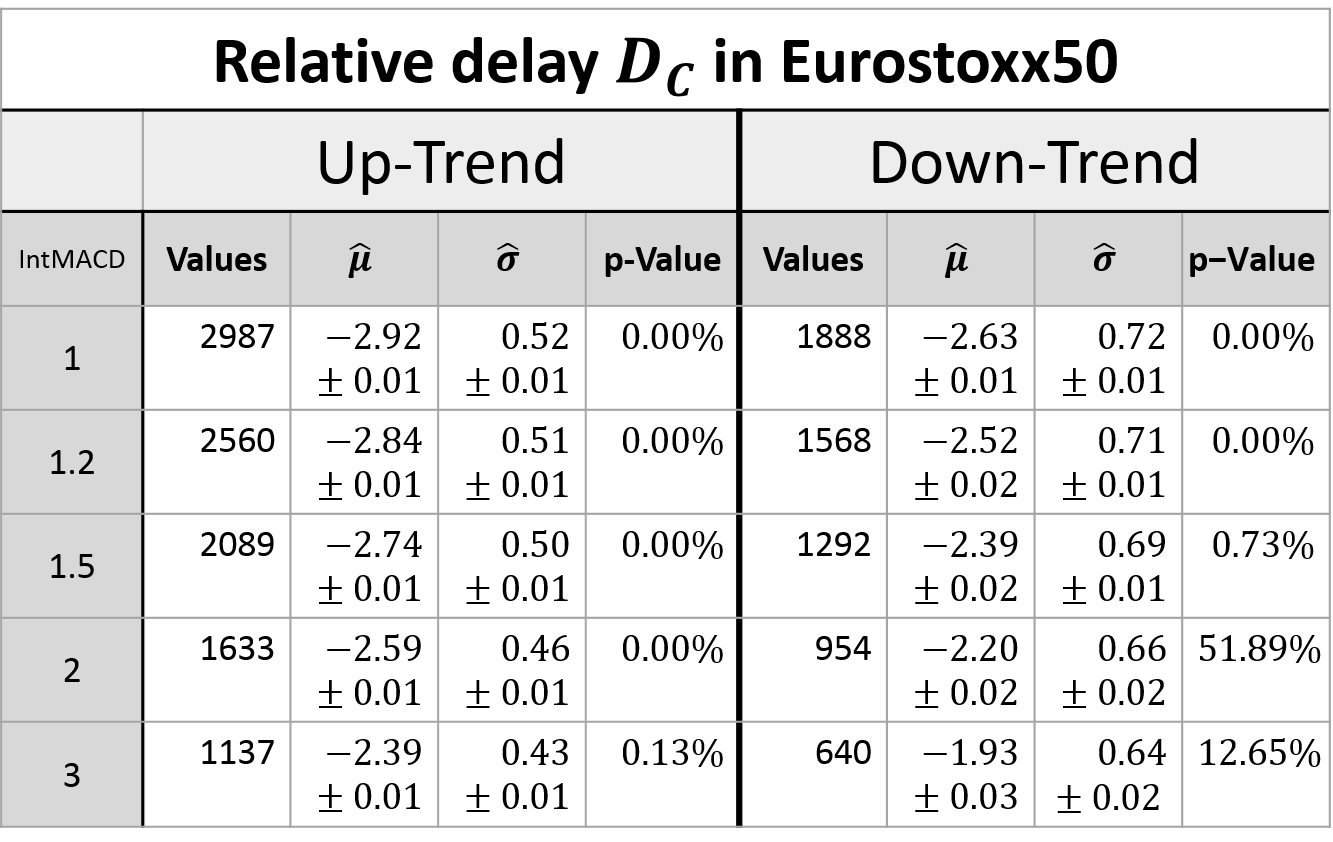}
  } \caption{Parameters of the log-normal fit for the relative delay $D_C$ after
  a correction in up- and down-trends.}
\label{table:kor_delay}
\end{table}
\begin{table}[p]
\centering
  \subfigure[$S\&P100$
  data]{\includegraphics[keepaspectratio=true,width=0.3\linewidth]{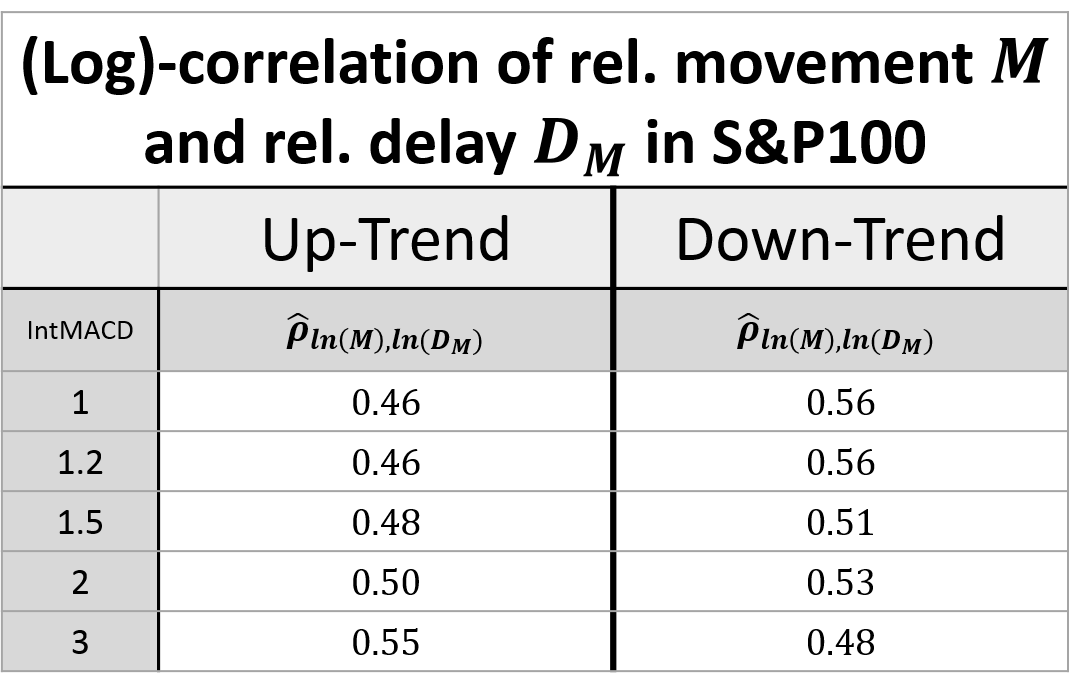}
  } \subfigure[$Eurostoxx50$
  data]{\includegraphics[keepaspectratio=true,width=0.3\linewidth]{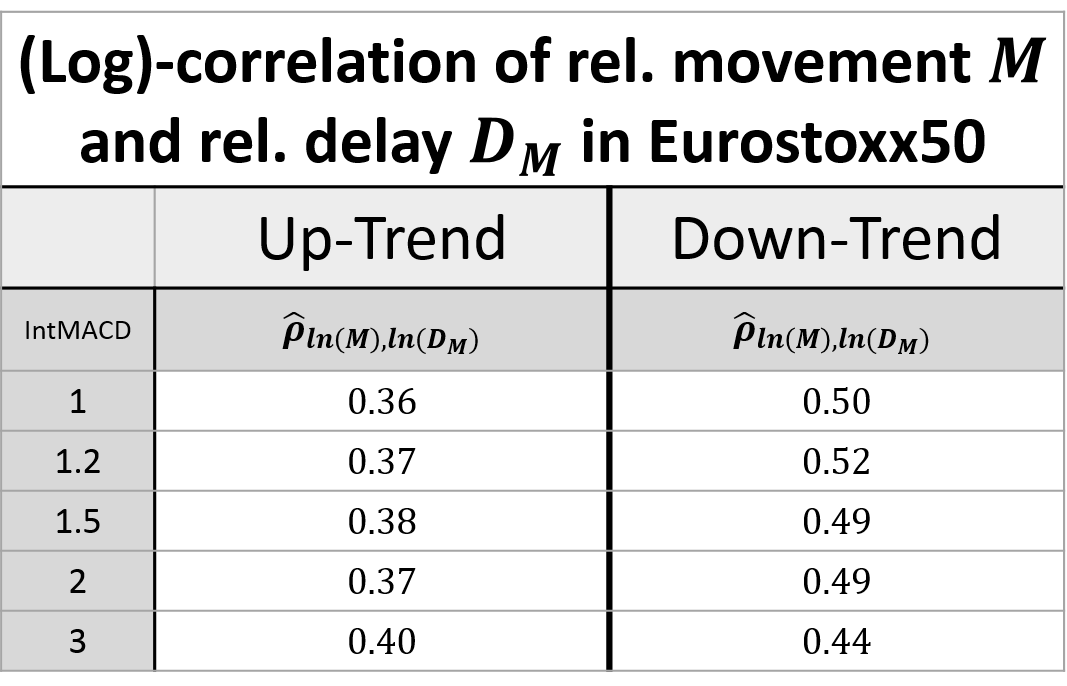}
  } \caption{Correlation between the logarithms of the relative movement $M$ and relative delay $D_M$ in up- and down-trends.}
\label{table:cor_bew}
\end{table}

\begin{table}[p]
\centering
  \subfigure[$S\&P100$
  data]{\includegraphics[keepaspectratio=true,width=0.3\linewidth]{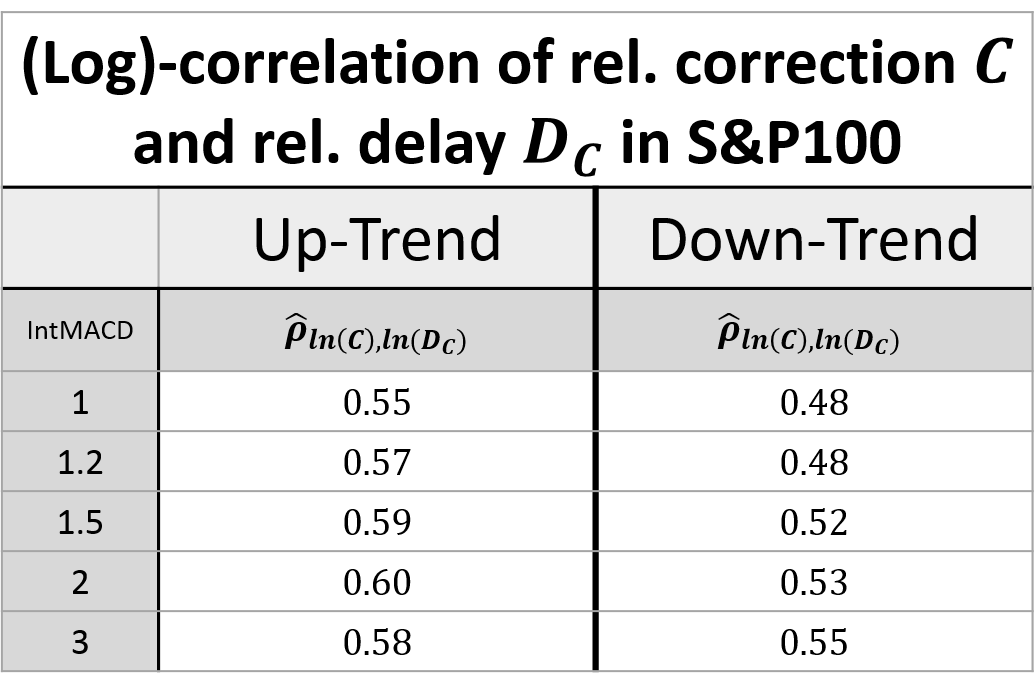}
  } \subfigure[$Eurostoxx50$
  data]{\includegraphics[keepaspectratio=true,width=0.3\linewidth]{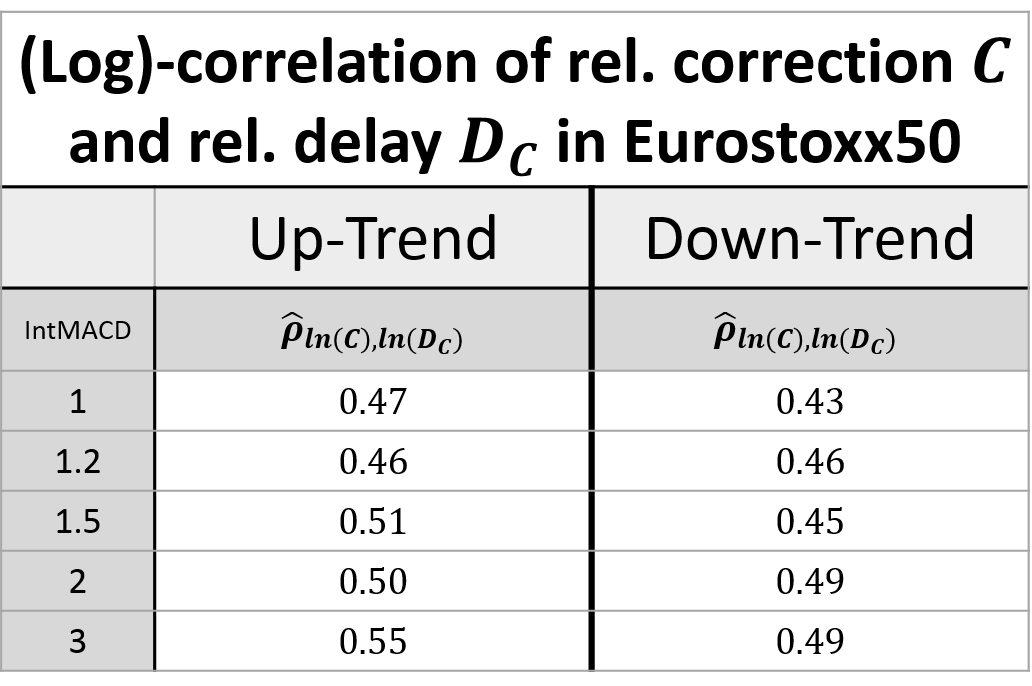}
  } \caption{Correlation between the logarithms of the relative correction $C$ and relative delay $D_C$ in up- and down-trends.}
\label{table:cor_kor}
\end{table}
The evaluation results for the relative delay are shown in Tables
\ref{table:bew_delay} to \ref{table:cor_kor}. It reveals the same behavior
of the model parameters as already seen for the relative movement and
correction. Furthermore, the knowledge of the correlations (Tables
\ref{table:cor_bew} and \ref{table:cor_kor}) enables joint considerations of the
relative movement/correction and the relative delay with the joint
distribution (\ref{eqn_xd}). In sum, this leads to the following expansion
of Observation \ref{beob_bew}.
\begin{Observation}[Log-Normal model for the rel.
movement/correction with delay]\label{beob_bew_delay}
The parameters of the log-normal
distribution $\mu$ and $\sigma$ are market invariant for the relative movement
and correction as well as their corresponding relative delays. Additionally, for
these trend variables the $\sigma$ parameter is more or less scale invariant
while $\mu$ increases for increasing scaling.\par
The parameters $\mu$ and $\sigma$ are also more or less
trend direction invariant for the relative movement (Table \ref{table:bew}). In
case of the relative correction, however, the direction of a trend affects these
parameters: Both are larger in case of a down-trend (Table \ref{table:kor}).
This is also true for the relative delay after a correction (Table
\ref{table:kor_delay}). For the relative delay after a movement $\mu$ is also
larger whereas $\sigma$ is smaller for down-trends (Table
\ref{table:bew_delay}).
\par Finally, the correlation between the logarithms of the relative movement/correction and the corresponding relative delay is close to scale invariant.
\end{Observation}
\subsection{Period of Movements and Corrections}
\begin{figure}[h]
  \centering
	\subfigure[Period $T$ for up-trends]
	{\includegraphics[keepaspectratio=true,width=0.55\linewidth]{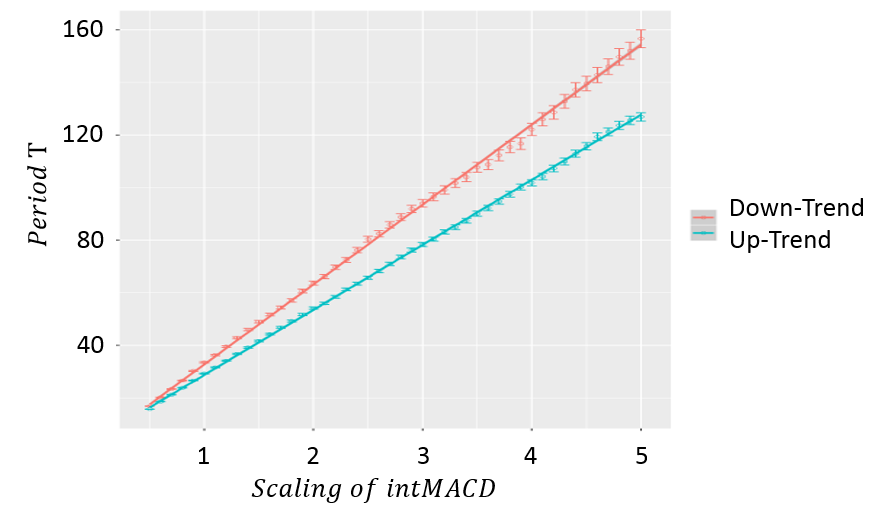}}
	\subfigure[Fit parameters of $T$]
	{\includegraphics[keepaspectratio=true,width=0.25\linewidth]{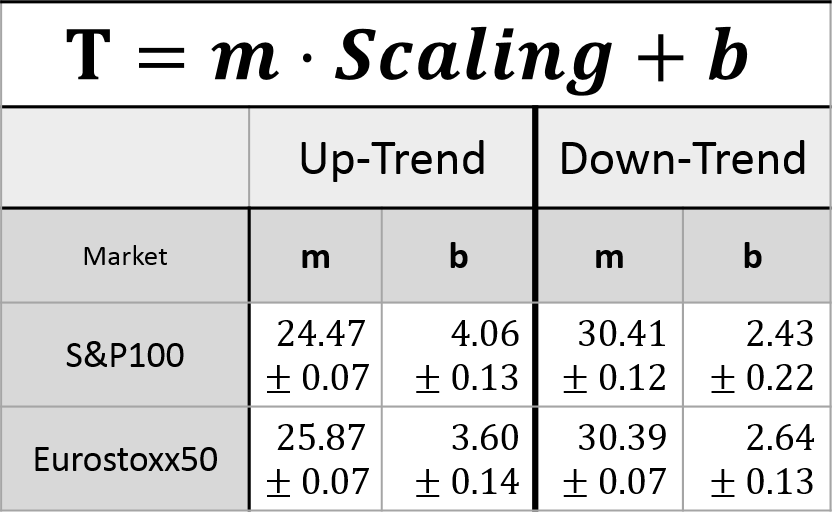}
	} \caption{Evolution of the period $T$ for scalings between $0.5$ and $5$ with
	a step size of $0.1$ for $S\&P100$ stocks.}
\label{fig:lambda_scaling}
\end{figure}

The dependency between $\mu$ and the scaling parameter has already been
explained with their connection to the level of trend significance (Observation
\ref{beob_bew}). One attribute of trend significance is the duration of a single
trend period, hence the time difference between two lows and two highs within the
up- and down-trend, respectively. 
It is called the \textit{period} $T$ of a trend.
Figure \ref{fig:lambda_scaling}.(a) shows the evolution of $T$ regarding
different scaling parameters. Here, for any scaling the $T$ value is the
arithmetical mean of all time differences between two consecutive lows and highs
within up- and down-trends, respectively.
The period $T$ shows a linear behavior which has already been observed in
\cite{EMP123} for EUR-USD Charts.
The fit parameters are similar for both evaluated markets but differ in regard to the type of trend (see Figure \ref{fig:lambda_scaling}.(b)). Due to
the linear model it is evident how to set the scaling parameter to emphasize a
specific period. Consequently, it is also easy to map any of the
three different trend classes introduced by Dow -- namely the primary, secondary and tertiary trend (see Murphy, chap. ``Dow Theory'' in \cite{MURPHY}).
\newpage
\section{Mathematical Model of Trading Systems}\label{sec:4}
\begin{figure}[h]
  \centering
  \subfigure[Parameters
  $a$ and $t$ with exemplary realizations of the retracement $X$
and delay $d$.]{
	\includegraphics[keepaspectratio=true,width=0.48\linewidth]{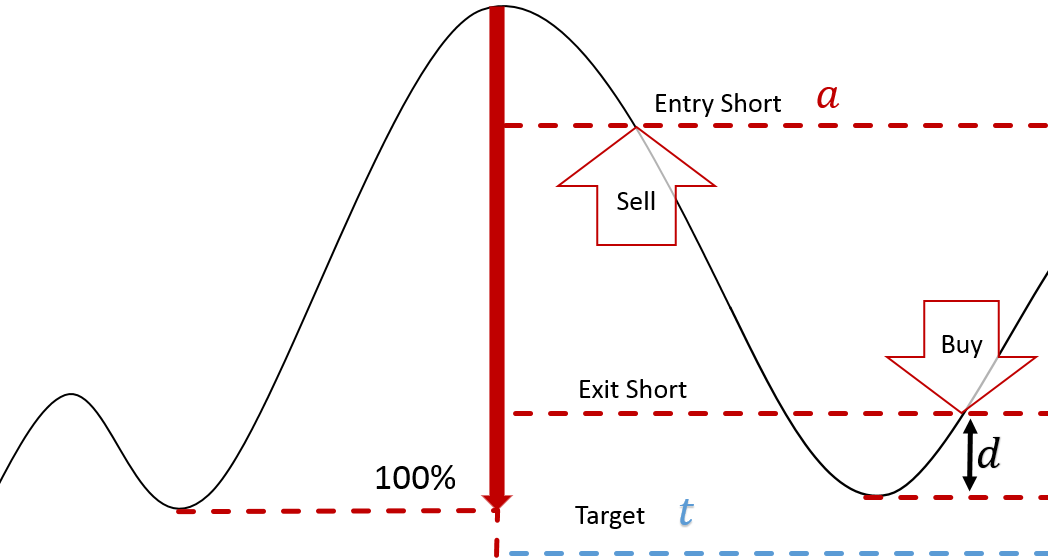}
	}
\subfigure[Entry at trend begin and following stop at last low.]{
\includegraphics[keepaspectratio=true,width=0.48\linewidth]{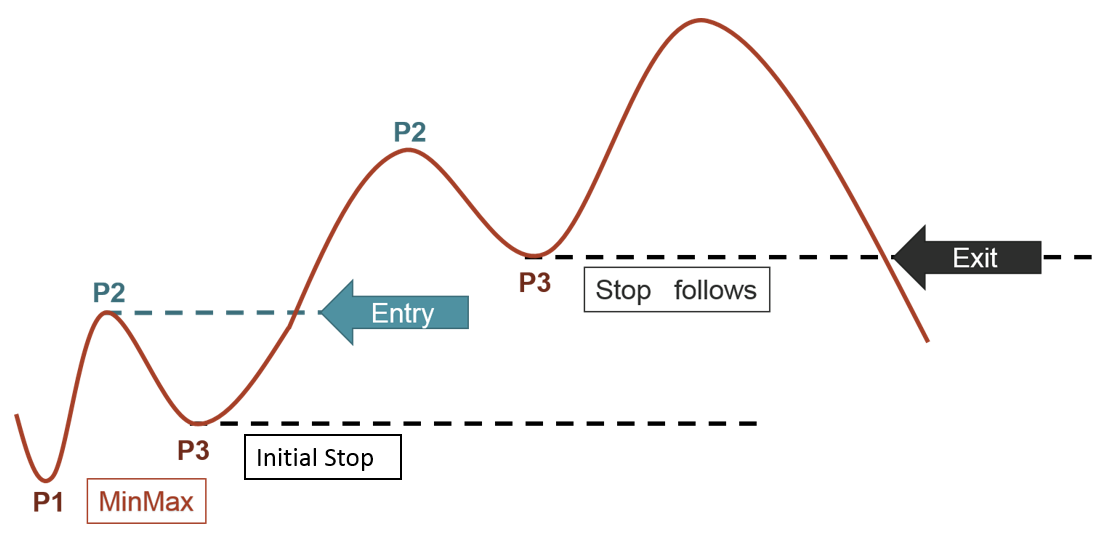}
}
\caption{Setup of a basic anti and pro cyclic trading system for up-trends (left
and right respectively).}\label{fig:hs}
\end{figure}

Based on the log-normal distribution model of the retracement an anti cyclic
trading system can be modeled for instance. With the joint density of
the retracement and delay the return of a basic anti cyclic trading system as shown
in Figure \ref{fig:hs} can be calculated:
\begin{Lemma}[Expected return]
Let an anti cyclic trading system as shown in Figure \ref{fig:hs}.(a) with entry
in the correction at retracement level $a$, target $t$ be given. As soon as the end
of the correction is recognized the position is closed with delay $d$.
Furthermore, the return (in units of the last movement) for a trade  with retracement $x$ and delay $d$ denoted by $R(x,d)$ is given by
\begin{equation*}
R(x,d)=
\begin{cases} 
x-a-d,\text{ if }a\leq x<t \quad\text{(retracement does not reach target)}\\
t-a,\text{ if }x\geq t \quad\text{(retracement reaches target)}
\end{cases}.
\end{equation*}
Moreover, the distribution of the random variable for the retracement $X$ and
for the delay after the the retracement $D=D_X$ are known from section
\ref{sec:2}.
That is the reason why, the expected value of this return, considering only retracements where
the trade is opened (condition $X\geq a$), is given by
\begin{eqnarray}
\E(R(X,D)|X\geq a)=&\,&\E(X|X\geq
a)-(a+\E(D|X\geq
a))\\
&+&\frac{1-F_X(t)}{1-F_X(a)}\left[t+\E(D|X\geq
t)-\E(X|X\geq
t)\right]\nonumber
\end{eqnarray}
with $F_X(x)=\Pe(X\leq x)$ denoting the distribution function of the retracement
$X$ (see section \ref{sec:2}).\par
\end{Lemma}
It should be noted, that $a$ and $t$ are parameters which have to be given in
the same units as the retracement, i. e. units of the last movement.
\begin{proof}
For a proof see \cite{MT_Kempen}.
\end{proof}
In the same manner, several other key figures such as variance of the return can be
calculated analytically assuming the empirically observed distributions as real. In this way,
a bad chance to risk ratio for the anti cyclic trading model has been revealed
(see \cite{MT_Kempen}). This is, however, in accordance with empirical
observations of backtests of that strategy.
\newpage
\section{Conclusion and Outlook}
In this survey the applications of the log-normal
distribution model on market-technical trend data is introduced. On the
one hand, it is remarkable that the log-normal model obviously fits better to
the trend data presented here than to daily returns of stock prices. In contrast
to the approach to daily returns of stock prices, however, there has not been
found an explanation for this observation yet. In particular, it has not yet been clarified whether the
log-normal distribution is a result of a limit process or can be explained with
the log-normal model for the daily returns of stock prices. As far as
applications in the direction of modeling of trading systems are concerned, we
introduced a simple model for an anti cyclic trading setup based on log-normally
distributed data.\par On the other hand, trend following, i.e.
pro cyclic trading systems are more widely used than anti cyclic ones. In fact, empirical backtests have already
shown the profitability of such trading systems. Consequently, there is a need
for a mathematical model. Unfortunately, pro cyclic trading usually implies
holding a position over several iterations of movement and correction as
outlined in Figure \ref{fig:hs}.(b).
This makes the problem far more complicated in mathematical terms since the joint
distribution of a random number of relative movements and corrections -- with
possible correlations -- has to be considered. Nevertheless, the log-normal
model for the trend data represents a promising approach to this issue as well.

\newpage
\renewcommand\bibname{References}

\end{document}